\documentclass[12pt]{article}
\usepackage{amsmath}
\usepackage{graphicx}
\usepackage{amsfonts}
\usepackage{amssymb}
\begin{document}

\title{An Introduction to Generalized Yang-Mills Theories}
\author{M. Chaves\\\textit{Universidad de Costa Rica, Escuela de Fisica}\\\textit{San Jose, Costa Rica}\\\textit{E-mail: mchaves@hermes.efis.ucr.ac.cr}}
\date{}
\maketitle
\begin{abstract}
In generalized Yang-Mills theories scalar fields can be gauged just as vector
fields in a usual Yang-Mills theory, albeit it is done in the spinorial
representation. The presentation of these theories is aesthetic in the
following sense: A physical theory using Yang-Mills theories requires several
terms and irreducible representations, but with generalized Yang-Mills
theories, only two terms and two irreducible representations are required.
These theories are constructed based upon the maximal subgroups of the gauge
Lie group. The two terms of the lagrangian are the kinetic energy of fermions
and of bosons. A brief review of Yang-Mills theories and covariant derivatives
is given, then generalized Yang-Mills theories are defined through a
generalization of the covariant derivative. Two examples are given, one
pertaining the Glashow-Weinberg-Salam model and another $SU(5)$ grand
unification. The first is based upon a $U(3)\supset U(1)\times U(1)\times
SU(2)$ generalized Yang-Mills theory, and the second upon a $SU(6)\supset
U(1)\times SU(5)$ theory. The possibility of expressing generalized Yang-Mills
theories using a five-dimensional formalism is also studied. The situation is
unclear in this case. At the end a list of comments and criticisms is given.
\end{abstract}
\tableofcontents

\section{Introduction.}

Yang-Mills theories have enjoyed tremendous success in the understanding and
quantization of three of the fundamental forces of nature: the
electromagnetic, weak and strong forces. The discovery and application of this
powerful idea was certainly one of the great scientific achievements of the
twentieth century. Their application to the field of elementary particles has
been so successful it has become the Standard Model of the field. Through
these theories the weak and electromagnetic force have been unified into a
renormalizable, phenomenologically correct theory based on the $SU(2)\times
U(1)$ Lie group. Quantum chromodynamics, based on $SU(3),$ has become by far
the most promising theory we have to understand the nonperturbative
complexities of the strong force. Finally, the efforts that have been done at
grand unification, although unsatisfactory in last analysis, have been
successful enough to show that, whatever form the final theory we all wait for
may have, at low energies it is going to look very much like a Yang-Mills theory.

In spite of all these achievements there is a very understandable attitude of
dissatisfaction in the overall status of the Standard Model. The large number
of empirical parameters, the arbitrary structure of the Higgs sector, the
rather \emph{ad hoc} multiplicity of terms and irreducible representations
(irreps), plus several puzzling situations it presents, such as the hierarchy
and generation problems, are all unpleasant reminders that we need a more
powerful and general theory to be able satisfactorily understand the workings
of the universe.

Here we give an introduction to Generalized Yang-Mills Theories (GYMTs),
which, as their name implies, are a generalization of the usual Yang-Mills
theories (YMTs). As it is the case in YMTs, a local gauge symmetry is enforced
upon the lagrangian; however, the role of gauge fields is not taken only by
vector fields, but also by scalar ones. The resulting theory is still gauge
invariant, but it allows the Higgs fields of quantum field theories to be
included as part of the covariant derivative.

These theories have not been properly structured yet as mathematical theories.
Their development till now has been from a practical point of view, keeping in
mind only their empirical applicability to the theories of high energy
physics. They have been applied successfully to two particular examples: the
Glashow-Weinberg-Salam Model (GWSM), and a case of grand unification. In both
cases the Higgs fields have been incorporated into the covariant derivative,
the many terms of each of these theories reduced to only two, and the large
number of irreps of each of these theories reduced to only two: the irrep of
the fermions and the irrep of the bosons (always the adjoint).

For each Lie group and one of its maximal subgroups there seems to be a GYMT.
The GWSM's GYMT is based upon $SU(3)\supset SU(2)\times U(1)$, and $SU(5)$
grand unification upon $SU(6)\supset SU(5)\times U(1)$.

We shall be using the metric
\begin{equation}
(\eta_{\mu\nu})=\mathrm{diag}(1,-1,-1,-1), \label{metric}%
\end{equation}
and the Dirac matrices $\gamma^{\mu},$ $\mu=0,1,2,3,$ obeying
\[
\gamma^{\mu}\gamma^{\nu}+\gamma^{\nu}\gamma^{\mu}=2g^{\mu\nu}.
\]
We use a representation in which $\gamma^{0\dagger}=\gamma^{0}$ and
$\gamma^{i\dagger}=-\gamma^{i},$ $i=1,2,3.$ We also use the chirality matrix
$\gamma^{5}=i\gamma^{0}\gamma^{1}\gamma^{2}\gamma^{3},$ so that $\gamma
^{5\dagger}=\gamma^{5}.$

\section{Brief review of Yang-Mills theories.}

The notion of gauge invariance goes back to Weyl$^{1}$. Consider the quantum
electrodynamics lagrangian
\begin{equation}
\mathcal{L}_{QED}=\bar{\psi}(i\partial\!\!\!/-eA\!\!\!/)\psi-\frac{1}{4}%
F^{\mu\nu}F_{\mu\nu} \label{QED}%
\end{equation}
where $F_{\mu\nu}\equiv\partial_{\mu}A_{\nu}-\partial_{\nu}A_{\mu}.$ This
lagrangian has a straightforward gauge invariance in the sense that the
Maxwell tensor $F_{\mu\nu},$ is invariant under the substitution $A_{\mu
}\rightarrow A_{\mu}+\partial_{\mu}\Lambda.$ The electron wavefunction is
invariant under a global (that is, spacetime independent) transformation
$\psi\rightarrow e^{i\delta}\psi$. To make it invariant under a local
transformation we must perform a simultaneous transformation on the
electromagnetic potential field $A_{\mu}$ and the electron wavefunction
$\psi,$ otherwise there would be an extra term left from the application of
Leibnitz's differentiation rule to the product $e^{i\delta}\psi$.

From a mathematical point of view the electromagnetic potential is a
connection dictating how vector fields are going to displace when they move
along specific paths along the base manifold $R^{3+1}.$ Its gauge invariance
allows for the establishment of a principal vector bundle structure, with
different connections, that is, electromagnetic potentials, over each open set
of the covering of the base space. In the overlapping zones of two open sets
the respective connections must only differ by a gauge transformation
$\partial_{\mu}\Lambda$. This conditions ensures that parallel transport
develops smoothly as any path is traversed along the base manifold. It is
interesting that from a mathematical point of view gauge freedom has the
advantage of allowing topologically nontrivial vector bundles, while from a
physical point of view the main advantage of gauge freedom is that it allows
for an improvement of the divergence of momentum integrals in quantum loop calculations.

\subsection{An abelian Yang-Mills theory.}

Let $U=e^{-i\alpha(x)}$ be an element of a local transformation based on the
$U(1)$ Lie group, so that the electrically charged fermion field transforms as
$\psi\rightarrow U\psi.$ The electromagnetic potential is required to obey the
gauge transformation law
\begin{equation}
A_{\mu}\rightarrow A_{\mu}+e^{-1}(\partial_{\mu}\alpha)\,{,} \label{gauge1}%
\end{equation}
so that, defining
\begin{equation}
D_{\mu}\equiv\partial_{\mu}+ieA_{\mu}, \label{covariant}%
\end{equation}
the covariant derivative must transform as
\begin{equation}
D_{\mu}\rightarrow UD_{\mu}U^{-1}\,{.} \label{transforms}%
\end{equation}
In this equation the derivative that is part of the covariant derivative is
acting indefinitely to the right and not only on the $U^{-1}$. We shall call
such operators \emph{unrestrained,} while the ones that act only on the
immediately following object, such as the partial in (\ref{gauge1}), we shall
call \emph{restrained}, and use a parenthesis to emphasize that the action of
the differentiation operator does not extend any further to the right\emph{.}
Note that while the operators $D_{\mu}$ and $D_{\mu}D_{\nu}$ are unrestrained,
the operator $[D_{\mu},D_{\nu}]$ is restrained, even though it \emph{looks} unrestrained.

Let us verify it really is restrained. Let $f=f(x)$ be a twice differentiable
function. Then
\begin{align}
-ie^{-1}[D_{\mu},D_{\nu}]f  &  =\partial_{\mu}A_{\nu}f-A_{\nu}\partial_{\mu
}f-\partial_{\nu}A_{\mu}f+A_{\mu}\partial_{\nu}f\nonumber\\
&  =\left(  (\partial_{\mu}A_{\nu})-(\partial_{\nu}A_{\mu})\right)  f,
\label{Leibnitz}%
\end{align}
an expression that has no unrestrained derivatives. There are mathematical and
physical reasons why the curvature, which, as we shall see, is what the
expression $[D_{\mu},D_{\nu}]$ represents, should contain only restrained
differential operators. In any case the point we want to make here is that in
GYMT all expressions of physical significance always automatically turn out to
have restrained operators.

Let us rewrite the lagrangian using the covariant derivative:
\begin{equation}
\mathcal{L}_{QED}=\bar{\psi}iD\!\!\!\!/\psi+\frac{1}{4e^{2}}[D_{\mu},D_{\nu
}][D^{\mu},D^{\nu}]. \label{QEDD}%
\end{equation}
The advantages with this formulation are twofold: on one hand the mathematical
meaning is transparent, and on the other the proof of gauge invariance becomes
straightforward. Thus, the effect of the gauge transformation on $\bar{\psi
}D\!\!\!\!/\psi$ results in $\bar{\psi}D\!\!\!\!/U^{-1}U\psi,$ but, since the
derivative is acting on all that follows to its right, the unitary operators
cancel and invariance follows. To verify the invariance of the second term on
the right of (\ref{QEDD}) notice that under the gauge transformation
(\ref{transforms}),
\[
\lbrack D_{\mu},D_{\nu}][D^{\mu},D^{\nu}]\rightarrow U[D_{\mu},D_{\nu}%
]U^{-1}U[D^{\mu},D^{\nu}]U^{-1}.
\]
Since we have already verified that the derivatives in the commutators of
$D$'s do not act on anything that comes after them, the unitary operators can
be commuted with the commutators and they cancel, leaving the original expression.

\subsection{A nonabelian Yang-Mills theory.}

The gauge structure of electromagnetism was generalized by Yang and
Mills$^{2}.$ The fermions transform as before $\psi\rightarrow U\psi,$ but in
their generalization of electromagnetism the $U$ do not commute among
themselves. The gauge fields are contracted with matrices that are a
representation of the adjoint representation of simple Lie groups. The
dimensionality of the adjoint representation is equal to the number of
generators of the Lie group. One associates to each generator one vector
field, and the sum of the products of each generator times a vector field
constitutes the connection.

To illustrate a nonabelian YMT let us take the Lie Group to be $SU(N).$ Its
fundamental representation $\mathbf{N}$ is of dimension $N,$ and its adjoint
$\mathbf{Adj}$ is of dimension $N^{2}-1.$ Let $A_{\mu}^{a}$, $a=1,\ldots
,N^{2}-1,$ be the fields to be associated with the generators. We take the
fermions to be in the fundamental representation $\mathbf{N,}$ so that the
wavefunction $\psi(x)$ has $N$ spinorial components. However, in order to
construct the lagrangian, we do not work with the $(N^{2}-1)\times(N^{2}-1)$
matrices that constitute the usual representation of the adjoint. The way to
proceed instead is to use the fact that for $SU(N),$ $\mathbf{N\times N}%
^{\ast}=\mathbf{Adj+1.}$ This implies that there must exist matrices $T^{a}$
of size $N\times N$ that are the Clebsch-Gordan coefficients generators
relating $\mathbf{N,}$ $\mathbf{N}^{\ast}$ and $\mathbf{Adj.}$ As a matter of
fact these Clebsch-Gordan coefficients in this case are no other than the
generators in the fundamental representation. Thus for these groups $A_{\mu
}^{a}T^{a}$ is a $SU(N)$ tensor transforming as
\begin{equation}
A_{\mu}^{a}T^{a}\rightarrow UA_{\mu}^{a}T^{a}U^{-1}, \label{tensor}%
\end{equation}
where $U$ is an element of the fundamental representation $\mathbf{N.}$ The
$A_{\mu}^{a},$ seen as an $N^{2}-1$ component vector, transforms in
(\ref{tensor}) as the $\mathbf{Adj}$ representation. The expression $\bar
{\psi}\gamma^{\mu}A_{\mu}^{a}T^{a}\psi$ is obviously gauge-invariant.

To assure gauge invariance in the presence of the fermion kinetic energy
derivative we must introduce a covariant derivative again. Thus we define
\begin{equation}
D_{\mu}=\partial_{\mu}+A_{\mu},\qquad A_{\mu}=igA_{\mu}^{a}(x)T^{a}
\label{nonabelian}%
\end{equation}
where $g$ is a coupling constant, and assign the following gauge
transformation to this dynamic field:
\begin{equation}
A_{\mu}\rightarrow UA_{\mu}U^{-1}-(\partial_{\mu}U)U^{-1}. \label{gauge2}%
\end{equation}
This in turn insures that the covariant derivative transforms similarly to the
way it did in the abelian case, that is, as in (\ref{transforms}).

The lagrangian of the YMT is then:
\begin{align}
\mathcal{L}_{YMT}  &  =\bar{\psi}iD\!\!\!\!/\psi+\frac{1}{2g^{2}}%
\widetilde{\operatorname*{Tr}}\left(  [D_{\mu},D_{\nu}][D^{\mu},D^{\nu
}]\right) \nonumber\\
&  =\bar{\psi}i(\partial\!\!\!/+A\!\!\!/)\psi+\frac{1}{2g^{2}}\widetilde
{\mathrm{Tr}}\left(  \partial_{\lbrack\mu}A_{\nu]}+[A_{\mu},A_{\nu}]\right)
^{2}, \label{YMT}%
\end{align}
where the trace is over the $SU(N)$ group generators and we follow the usual
normalization condition%
\begin{equation}
\widetilde{\mathrm{Tr}}T^{a}T^{b}=\frac{1}{2}\delta_{ab}. \label{normhalf}%
\end{equation}
A tilde is used over the trace to differentiate it from the trace over Dirac
matrices that we shall also use. We are missing in this lagrangian scalar
boson Higgs fields. With them it is possible to construct other terms in the
Lagrangian that are also simultaneously Lorentz- and gauge-invariant and are
of the general form $\overline{fermion}\times Higgs\times fermion$. These,
called the Yukawa terms, are of extreme importance since they constitute the
natural channel these theories have to generate fermion mass. This comes about
through a hypothesized nonzero vacuum expected value (VEV) of the Higgs. The
lagrangian must also contain the kinetic energy of the Higgs, that has the
form $|D_{\mu}\varphi|^{2}.$ Furthermore, to justify the nonzero VEV for the
Higgs another term is usually added, a potential $V(\varphi)$ that has a local
minimum at some constant and uniform value $\varphi=v.$

\section{The gauge field as a connection.}

The local antihermitian matrix function defined in (\ref{nonabelian}) can
serve as a connection. Let $\phi(x)$ be an element at point $x$ of an $N$
dimensional vector space. Then, for an infinitesimal path $dx$ in $R^{1+3}$ we
\emph{define} the parallel displacement of $\phi$ to be
\begin{equation}
d\phi_{||}\equiv dx^{\mu}A_{\mu}\phi. \label{connection}%
\end{equation}
Notice that the vector does not change length in a YMT due to its parallel
displacement. The change in $\phi$'s length is given by
\[
(\phi+d\phi_{||})^{\dagger}(\phi+d\phi_{||})-\phi^{\dagger}\phi\approx
dx^{\mu}\phi^{\dagger}(A_{\mu}+A_{\mu}^{\dagger})\phi=0
\]
due to the antihermitian nature of $A_{\mu}.$

\subsection{An unitary operator as functional of a path.}

Consider now a path $C$ in $R^{1+3}$ from $x_{1}$ to $x_{2}$ parametrized with
$s,$ $0\leq s\leq1,$ so that $x$ is a function $x(s)$ and $x(0)=x_{1},$
$x(1)=x_{2}.$ Since parallel displacement does not change the length of the
field $\phi$ its effect can be simply expressed as multiplication by a unitary
operator $U(x(s)).$ The change in the vector at any particular point $s$ in
$C$ is given by $dU\phi=dx^{\mu}A_{\mu}U\phi,$ so the unitary operator has to
satisfy the equation
\begin{equation}
\frac{dU}{ds}=\frac{dx^{\mu}}{ds}A_{\mu}U. \label{differential}%
\end{equation}
Its solution is
\begin{equation}
U(s)=Pe^{I(s)},\qquad I=\int_{0}^{s}ds\frac{dx^{\mu}}{ds}A_{\mu},
\label{solution}%
\end{equation}
where $P$ is the path ordering operator, which arranges operators so that they
are placed according to the value of $s$ in their argument, from higher values
on the left to lower on the right. Let us prove that (\ref{solution}) is
really the solution of the differential equation. The following is a simple
demonstration that does not involve changing the integration limits$:$%
\begin{align}
\frac{dU}{ds}  &  =\frac{d}{ds}P\left(  1+I+\frac{1}{2}I^{2}+\frac{1}{3!}%
I^{3}+\ldots\right) \nonumber\\
&  =\dot{I}+\frac{1}{2}P(\dot{I}I+I\dot{I})+\frac{1}{3!}P(\dot{I}II+I\dot
{I}I+II\dot{I})+\ldots\nonumber\\
&  =\dot{I}P\left(  1+I+\frac{1}{2}I^{2}+\ldots\right)  =\frac{dx^{\mu}}%
{ds}A_{\mu}U. \label{wilson}%
\end{align}
The path ordering operator $P$ does not affect the unitarity of the $U(s).$
One can show that $U^{\dagger}=U^{-1}$ simply by expanding the series and
performing the required algebra. This is consistent with our previous result
that parallel transport cannot change the length of a vector.

\subsection{Group generated by integration over closed paths.}

Consider the set $\{C_{Q}\}$ of all continuous closed paths in spacetime that
pass through the point $Q,$ and let $U(C_{Q})=P\exp\oint_{C_{Q}}dx^{\mu}%
A_{\mu}.$ Then the set $G_{Q}=\{U(C_{Q})\}$ of unitary operators for all such
paths forms a group. The identity corresponds to a path made up of the single
point $Q.$ The inverse element to an operator is one with the same integration
path but traversed instead in the inverse direction. The path associated with
the product of two operators with paths $C_{Q}$ and $C_{Q}^{\prime}$ is one
that has a single path $C_{Q}^{\prime\prime}=C_{Q}+C_{Q}^{\prime}$ defined as
follows: it begins at $Q$ and follows $C_{Q}$ till it comes back to $Q,$ then
joins $C_{Q}^{\prime}$ and follows it till it comes back to $Q,$ and then
joins $C_{Q}$ again, forming a closed curve. The closure of the product of two
operators hinges precisely on the presence of the path operator $P$. Thus let
two operators be $e^{I(C_{Q})}$ and $e^{I(C_{Q}^{\prime})},$ so that their
product should then be
\[
e^{I(C_{Q})}e^{I(C_{Q}^{\prime})}=e^{I(C_{Q}^{\prime\prime})}.
\]
But clearly this is going to be true only if $P[I(C_{Q}),I(C_{Q}^{\prime
})]=0.$ One effect of the path ordering operator is to order the integrands in
way that is unconnected to their original order, so that this commutator has
to be zero.

\subsection{Example of the meaning of curvature taken from Riemannian spaces.}

Let us introduce curvature in the context of Riemannian geometry. It is
well-known that geodesic deviation and parallel transport around an
infinitesimally small closed curve in a Riemann manifold are two aspects of
the same construction.$^{3}$ Consider a four dimensional Riemann manifold with
metric $g_{\mu\nu}$ and a vector $B^{\mu}$ whose parallel transport around a
closed curve $C$ is given by
\begin{equation}
\Delta B_{\alpha}=\oint_{C}\frac{dx^{\nu}}{ds}\Gamma^{\sigma}{}_{\nu\alpha
}B_{\sigma}\,ds, \label{path}%
\end{equation}
where $s$ is a parametrization of the curve and $\Gamma^{\sigma}{}_{\nu\alpha
}$ is the connection that acts on the vectors of the tangent vector space to
the Riemann manifold, the Christoffel symbol. For simplicity we take $C$ be a
small parallelogram made up of the two position vectors $x^{\mu}$ and $y^{\mu
},$ so that its area two-form is $\Delta S^{\mu\nu}\equiv\frac{1}{2}x^{[\mu
}y^{\nu]}.$ To perform the integral we first make, about the center of the
parallelogram, a Taylor expansion
\begin{equation}
\tilde{\Gamma}^{\sigma}{}_{\nu\alpha}=\Gamma^{\sigma}{}_{\nu\alpha}%
+\Gamma^{\sigma}{}_{\nu\alpha,\mu}\Delta x^{\mu} \label{tildechrist}%
\end{equation}
for the Christoffel symbol and a parallel displacement
\begin{equation}
\tilde{B}_{\sigma}=B_{\sigma}+\Delta x^{\mu}\Gamma^{\tau}{}_{\mu\sigma}%
B_{\tau} \label{tildevector}%
\end{equation}
for the vector, both ending at some point separated a distance $\Delta x^{\mu
}$ from the center. The quantities with tilde will take their values at each
of the four sides of the parallelogram, which we shall call \#1, \#2, \#3 and
\#4, counting counterclockwise. The contribution of side \#1 to the integral
is:
\[
\Delta y^{\nu}\tilde{\Gamma}^{\sigma}{}_{\nu\alpha}\tilde{B}_{\sigma}%
|_{1}=y^{\nu}(\Gamma^{\sigma}{}_{\nu\alpha}+\Gamma^{\sigma}{}_{\nu\alpha,\mu
}x^{\mu}/2)(B_{\sigma}+x^{\mu}\Gamma^{\tau}{}_{\mu\sigma}B_{\tau}/2).
\]
The contribution of side \#3 is
\[
\Delta y^{\nu}\tilde{\Gamma}^{\sigma}{}_{\nu\alpha}\tilde{B}_{\sigma}%
|_{3}=-y^{\nu}(\Gamma^{\sigma}{}_{\nu\alpha}-\Gamma^{\sigma}{}_{\nu\alpha,\mu
}x^{\mu}/2)(B_{\sigma}-x^{\mu}\Gamma^{\tau}{}_{\mu\sigma}B_{\tau}/2).
\]
The contribution of side \#1 and its opposite side \#3 is
\[
\Delta y^{\nu}\tilde{\Gamma}^{\sigma}{}_{\nu\alpha}\tilde{B}_{\sigma}%
|_{1+3}=x^{\mu}y^{\nu}(\Gamma^{\sigma}{}_{\nu\alpha}\Gamma^{\tau}{}_{\mu
\sigma}+\Gamma^{\tau}{}_{\nu\alpha,\mu})B_{\tau},
\]
up to first order. The contribution of side \#2 is:
\[
\Delta x^{\mu}\tilde{\Gamma}^{\sigma}{}_{\mu\alpha}\tilde{B}_{\sigma}%
|_{2}=-x^{\mu}(\Gamma^{\sigma}{}_{\mu\alpha}+\Gamma^{\sigma}{}_{\mu\alpha,\nu
}y^{\nu}/2)(B_{\sigma}+y^{\nu}\Gamma^{\tau}{}_{\nu\sigma}B_{\tau}/2),
\]
and that of its opposite side \#4 is
\[
\Delta x^{\mu}\tilde{\Gamma}^{\sigma}{}_{\mu\alpha}\tilde{B}_{\sigma}%
|_{4}=x^{\mu}(\Gamma^{\sigma}{}_{\mu\alpha}-\Gamma^{\sigma}{}_{\mu\alpha,\nu
}y^{\nu}/2)(B_{\sigma}-y^{\nu}\Gamma^{\tau}{}_{\nu\sigma}B_{\tau}/2),
\]
so that
\[
\Delta x^{\mu}\tilde{\Gamma}^{\sigma}{}_{\mu\alpha}\tilde{B}_{\sigma}%
|_{2+4}=-x^{\mu}y^{\nu}(\Gamma^{\sigma}{}_{\mu\alpha}\Gamma^{\tau}{}%
_{\nu\sigma}+\Gamma^{\tau}{}_{\mu\alpha,\nu})B_{\tau}.
\]
Summing over the four sides one obtains for the change of the vector as it is
parallel-transported around the parallelogram:
\begin{align}
\Delta B_{\alpha}  &  \equiv\oint_{C}\Delta x^{\mu}\tilde{\Gamma}^{\sigma}%
{}_{\mu\alpha}\tilde{B}_{\sigma}\nonumber\\
&  =x^{\mu}y^{\nu}(\Gamma^{\tau}{}_{\nu\alpha,\mu}-\Gamma^{\tau}{}_{\mu
\alpha,\nu}+\Gamma^{\sigma}{}_{\nu\alpha}\Gamma^{\tau}{}_{\mu\sigma}%
-\Gamma^{\sigma}{}_{\mu\alpha}\Gamma^{\tau}{}_{\nu\sigma})B_{\tau}\nonumber\\
&  =x^{\mu}y^{\nu}R^{\tau}{}_{\alpha\mu\nu}B_{\tau}=\Delta S^{\mu\nu}R^{\tau
}{}_{\alpha\mu\nu}B_{\tau}, \label{deficit}%
\end{align}
where $R^{\tau}{}_{\alpha\mu\nu}$ is the Riemann curvature tensor and we have
used the antisymmetry of its last two indices. Thus the change in $B_{\sigma}$
is proportional to the surface enclosed by the path. If the Riemann tensor is
zero, the change in $B_{\sigma}$ is null.

For convenience we list some properties this tensor satisfies:
\[%
\begin{tabular}
[c]{ll}%
$R^{\tau}{}_{\alpha\mu\nu}=-R^{\tau}{}_{\alpha\nu\mu},$ & $R^{\tau}{}%
_{[\alpha\mu\nu]}=0,$\\
$R^{\tau}{}_{\alpha\lbrack\mu\nu;\omega]}=0,$ & $R_{\alpha\beta\mu\nu
}=-R_{\beta\alpha\mu\nu},$\\
$R_{\alpha\beta\mu\nu}=R_{\mu\nu\alpha\beta},$ & $R_{[\alpha\beta\mu\nu]}=0.$%
\end{tabular}
\]

\subsection{Riemann curvature tensor as a commutator of covariant derivatives.}

We wish to write the curvature tensor as a commutator of covariant
derivatives, as we did with YMT. We are going to prove it in a way that
emphasizes the similarity between YMTs and the Theory of General Relativity
(TGR), because of the suggestiveness of the exercise.

The covariant derivative of a vector $B_{\sigma}$ is given by
\begin{equation}
B_{\sigma;\mu}=\partial_{\mu}B_{\sigma}-\Gamma^{\tau}{}_{\mu\sigma}B_{\tau
}=(\partial_{\mu}\delta^{\tau}{}_{\sigma}-\Gamma^{\tau}{}_{\mu\sigma})B_{\tau
}. \label{covariant derivative}%
\end{equation}
This is the standard formalism used in the TGR, and uses no implicit indices
at all; all are explicit. In YMTs the indices corresponding to internal space,
or, as mathematicians say it, the indices of the matrices that make up the
transition group and the connection, are left implicit with a remarkable
simplification resulting. Let us write (\ref{covariant derivative}) leaving
implicit the indices that have to do with the tangential vector space. We the
assume that $\delta^{\tau}{}_{\sigma}$ is the identity matrix $\mathbf{1}$
with implicit subindices, and that $\Gamma^{\tau}{}_{\mu\sigma}$ is one of a
set of four matrices $A_{\mu}$ also with implicit subindices. Then
\[
\delta^{\tau}{}_{\sigma}\rightarrow\mathbf{1}_{\sigma\tau},\qquad\Gamma^{\tau
}{}_{\mu\sigma}\rightarrow(A_{\mu})_{\sigma\tau},
\]
and the covariant derivative of (\ref{covariant derivative}) can be written
\[
B_{\sigma;\mu}\rightarrow(\mathbf{1}\partial_{\mu}-A_{\mu})B,
\]
where the index of the vector $B_{\sigma}$ is, naturally, also kept implicit.
In what follows we also omit the identity matrix $\mathbf{1,}$ as is almost
always done in similar occasions in physics. We can now achieve our goal very
efficiently. Notice, too, how unrestrained differentials never appear in the
TGR. The previous expression for the Riemann curvature tensor can be written
in the interesting form
\[
\lbrack D_{\mu},D_{\nu}]=\partial_{\lbrack\nu}A_{\mu]}+[A_{\mu},A_{\nu}].
\]
To prove this we re-introduce the indices and usual names of the variables in
the commutator that appears in above's equation. It immediately follows that
\begin{align*}
\lbrack D_{\mu},D_{\nu}]_{\sigma\tau}  &  =\Gamma^{\tau}{}_{\mu\sigma,\nu
}-\Gamma^{\tau}{}_{\nu\sigma,\mu}+\Gamma^{\beta}{}_{\mu\sigma}\Gamma^{\tau}%
{}_{\nu\beta}-\Gamma^{\beta}{}_{\nu\sigma}\Gamma^{\tau}{}_{\mu\beta}\\
&  =-R_{\sigma\mu\nu}^{\tau},
\end{align*}
that is, the commutator of the covariant derivative is minus the Riemann
curvature tensor.

The expressions in lagrangians (\ref{QEDD}) and (\ref{YMT}) contain terms that
strongly resemble the curvature expression of Riemannian spaces. As a matter
of fact they are the curvatures associated with the connections that appear in
principal vector bundles. Our interest here again is not to review general
physical and mathematical results, but to show the similarity of the
mathematical expressions in two different but equally fundamental physical
theories. There is one rather striking difference between these two theories:
in the YMT the curvature enters squared in the lagrangian, while in the TGR it
enters linear.

\subsection{Curvature in a Yang-Mills theory.}

We justify now our having identified $[D_{\mu},D_{\nu}]$ with the curvature.
Here $D_{\mu}$ is the Yang-Mills covariant derivative defined in
(\ref{nonabelian}). We proceed similarly to the way we did in the Riemannian
case. We define the change in a vector $\chi$ to be given by the parallel
transport of the vector around a closed path $C.$ We maintain the same
previous base manifold: four-dimensional spacetime with the flat metric
(\ref{metric}). The excess of the vector is:
\[
\Delta\chi=\oint\frac{dx^{\nu}}{ds}A_{\nu}\chi ds,
\]
where $s$ is a parametrization of the curve and $A_{\nu}$ is the Yang-Mills
connection. We can now mimic the derivation of the calculation we did of
parallel transport for in a Riemannian manifold with connection, and obtain
\begin{align*}
\Delta\chi &  =x^{\mu}y^{\nu}(A_{\nu,\mu}-A_{\mu,\nu}+A_{\nu}A_{\mu}-A_{\mu
}A_{\nu})\chi\\
&  =x^{\mu}y^{\nu}[D_{\mu},D_{\nu}]\chi\equiv\Delta S^{\mu\nu}F_{\mu\nu}\chi.
\end{align*}
We conclude that curvature in a YMT is given by
\[
F_{\mu\nu}=[D_{\mu},D_{\nu}].
\]

\section{Introduction to generalized Yang-Mills theories.}

The sense in which one generalizes a YMT is by promoting scalar fields to the
level of gauge fields. In the standard YMT, gauge invariance is assured by
demanding the vector gauge field transform as $A_{\mu}\rightarrow UA_{\mu
}U^{-1}-(\partial_{\mu}U)U^{-1}.$ A scalar gauge field does not have a vector
index to associate with the term $(\partial_{\mu}U)U^{-1}$ in the previous
equation. On way to include a one scalar field in a covariant derivative would
be to increase the dimension of the base manifold by one, so that
$\mu=0,1,2,3,4.$ However, there are two serious problems with this approach.
First, this is would not be a true generalization of YMTs, but simply a YMT
constructed in a higher dimensionality. Second, the resulting theory is
phenomenologically unacceptable: it does predict the physics of the standard model.

The way we proceed to be able to include scalar bosons as gauge fields

is to write a spinorial covariant derivative, and associate the scalar fields
with the chiral operator $\gamma^{5}.$ This procedure is very successful from
a phenomenological point of view. As usual the gauge bosons are placed in the
adjoint representation of a Lie group. The idea is that \textit{all} scalar
bosons have to enter the model this way. In this aspect, as in several others,
GYMTs are far more restrictive than the usual YMTs. The total number of
bosons, vector plus scalar, has to equal to the number of group generators,
since they have to be multiplied by the same $N^{2}-1$ matrices $T^{a}$ as we
saw in the YMTs. Fermions fields are placed in irreducible representations
(irreps) of mixed chirality according to rules given by the particular choice
of maximal subgroup of the original Lie group of the GYMT.

This setup presents us with a difficulty, since the kinetic energy of the
vector bosons is not written using a spinorial covariant derivative, so we
cannot use the generalized form of the covariant derivative we are planning to
implement. We address this difficulty next.

\subsection{Gauge field's kinetic energy using a spinorial covariant
derivative: abelian case.}

There is a $2\times1$ homomorphism between the Dirac spinorial and vector
representations of the Lorentz group. If $S$ is a matrix that represents an
element of the spinorial representation, and if $\psi$ is a spinor and
$A\!\!\!/$ a slashed 4-vector, then under a Lorentz transformation
\[
\psi\rightarrow S\psi,\qquad A\!\!\!/\rightarrow SA\!\!\!/S^{-1}.
\]
The following identity relates the vectorial formulation of the Yang-Mills
kinetic energy with the Dirac spinorial one.

\textbf{Theorem, abelian case. }\emph{Let }$D_{\mu}=\partial_{\mu}+B_{\mu}%
$\emph{, where }$B_{\mu}$\emph{\ is a vector field. Then: }
\begin{equation}
\left(  (\partial_{\mu}B_{\nu})-(\partial_{\nu}B_{\mu})\right)  \left(
(\partial^{\mu}B^{\nu})-(\partial^{\nu}B^{\mu})\right)  =\frac{1}%
{8}\mathrm{Tr}^{2}D\!\!\!\!/^{\,2}-\frac{1}{2}\mathrm{Tr}D\!\!\!\!/^{\,4},
\label{teo_1}%
\end{equation}
\emph{where the traces are to be taken over the Dirac matrices.}

The differentiations on the right of the equation are not restrained. To prove
the theorem it is convenient to use the differential operator
\[
O\equiv\partial^{2}+2B\cdot\partial+B^{2}.
\]
It is important to notice that it does not contain any contractions with Dirac
matrices, so that when it appears in traces of Dirac matrices it facilitates
the calculation. For example, $\mathrm{Tr}O=4O,$ $\mathrm{Tr}O\left(
\partial\!\!\!/B\!\!\!\!/\right)  =4O\left(  \partial\cdot B\right)  $, etc.
For convenience we give a list of Dirac trace formulae:
\begin{align}
\operatorname*{Tr}1  &  =4,\nonumber\\
\operatorname*{Tr}a\!\!\!/_{1}a\!\!\!/_{2}  &  =4a_{1}\cdot a_{2},\nonumber\\
\operatorname*{Tr}a\!\!\!/_{1}a\!\!\!/_{2}\cdots a\!\!\!/_{n}  &  =0,\text{ n
odd,}\nonumber\\
\operatorname*{Tr}\gamma^{5}a\!\!\!/_{1}a\!\!\!/_{2}\cdots a\!\!\!/_{n}  &
=0,\text{ n odd,}\nonumber\\
\operatorname*{Tr}\gamma^{5}  &  =0,\nonumber\\
\operatorname*{Tr}\gamma^{5}a\!\!\!/_{1}a\!\!\!/_{2}  &  =0,\label{list}\\
\operatorname*{Tr}a\!\!\!/_{1}a\!\!\!/_{2}a\!\!\!/_{3}a\!\!\!/_{4}  &
=4(a_{1}\cdot a_{2}\,a_{3}\cdot a_{4}-a_{1}\cdot a_{3}\,a_{2}\cdot a_{4}%
+a_{1}\cdot a_{4}\,a_{2}\cdot a_{3}).\nonumber
\end{align}

We now relate $O$ with $D\!\!\!\!/=\partial\!\!\!/+B\!\!\!\!/$ by taking the
square of this last quantity and letting it act on some twice differentiable
function $f=f(x)$:
\begin{align*}
D\!\!\!\!/D\!\!\!\!/f  &  =\partial^{2}f+\partial
\!\!\!/(B\!\!\!\!/f)+B\!\!\!\!/\,\partial\!\!\!/f+B^{2}f\\
&  =\partial^{2}f+\partial\!\!\!/(B\!\!\!\!/f)-\gamma^{\mu}%
B\!\!\!\!/\,\partial_{\mu}f+2B\cdot\partial+B^{2}f\\
&  =Of+\left(  \partial\!\!\!/B\!\!\!\!/\right)  f,
\end{align*}
or
\begin{equation}
D\!\!\!\!/D\!\!\!\!/=O+\left(  \partial\!\!\!/B\!\!\!\!/\right)  .
\label{trilobite}%
\end{equation}
We use this form of $D\!\!\!\!/^{\,2}$ in the traces on the right of
(\ref{teo_1}). The first trace takes the form
\begin{equation}
\frac{1}{8}\mathrm{Tr}^{2}\left[  O+\left(  \partial\!\!\!/B\!\!\!\!/\right)
\right]  =2\left(  O^{2}+O(\partial\cdot B)+(\partial\cdot B)O+(\partial\cdot
B)^{2}\right)  , \label{trace_1}%
\end{equation}
and the second
\begin{equation}
\frac{1}{2}\mathrm{Tr}\left[  O+\left(  \partial\!\!\!/B\!\!\!\!/\right)
\right]  ^{2}=2\left(  O^{2}+O(\partial\cdot B)+(\partial\cdot B)O\right)
+\frac{1}{2}\mathrm{Tr}\left[  \left(  \partial\!\!\!/B\!\!\!\!/\right)
\left(  \partial\!\!\!/B\!\!\!\!/\right)  \right]  . \label{trace_2}%
\end{equation}
Recalling the formula (\ref{list}) and using it for $a_{1}=\partial,$
$a_{2}=B,$ $a_{3}=\partial$ and $a_{4}=B,$ one obtains
\begin{equation}
\frac{1}{8}\mathrm{Tr}^{2}\left[  O+\left(  \partial\!\!\!/B\!\!\!\!/\right)
\right]  -\frac{1}{2}\mathrm{Tr}\left[  O+\left(  \partial
\!\!\!/B\!\!\!\!/\right)  \right]  ^{2}=(\partial_{\lbrack\mu}B_{\nu
]})(\partial^{\lbrack\mu}B^{\nu]}) \label{proof}%
\end{equation}
as we wished to demonstrate. There are two motivations for the including the
first trace term in (\ref{teo_1}): it ensures that the differential operators
be restrained, and it cancels some terms that look alien to high energy
quantum field theories.

With the help of the Theorem, we can write the second term of (\ref{QED}), the
QED Lagrangian, in the form
\begin{equation}
-\frac{1}{4}F^{\mu\nu}F_{\mu\nu}=e^{-2}\left(  \frac{1}{32}\mathrm{Tr}%
^{2}D\!\!\!\!/^{\,2}-\frac{1}{8}\mathrm{Tr}D\!\!\!\!/^{\,4}\right)  \,{.}
\label{QED'}%
\end{equation}
The point here is that we are using only the spinorial form of the covariant
derivative, so that we can generalize the derivative.

\subsection{Gauge field's kinetic energy using a spinorial covariant
derivative: nonabelian case.}

If we are dealing with a nonabelian YMT we have to revise the derivation of
the previous subsection looking for places where the new noncommutativity of
$B_{\mu}$ may make a difference. We begin with the same expression $\frac
{1}{8}\mathrm{Tr}^{2}D\!\!\!\!/^{\,2}-\frac{1}{2}\mathrm{Tr}D\!\!\!\!/^{\,4}$
and our interest is seeing if new terms have appeared in the right hand side
of (\ref{proof}).

Notice that in (\ref{trilobite}) it is no longer true that
$B\!\!\!\!/B\!\!\!\!/=B^{2},$ so that if we maintain the same definition for
$O$ the we must modify (\ref{trace_2}). On the other hand, (\ref{trace_1})
does not have to be changed, since it is still true that $\operatorname*{Tr}%
B\!\!\!\!/B\!\!\!\!/=4B^{2}.$ The new form of the second trace of the right
hand side of (\ref{teo_1}) is
\begin{equation}
\frac{1}{2}\mathrm{Tr}D\!\!\!\!/^{\,4}=\frac{1}{2}\mathrm{Tr}\left[
O-B^{2}+B\!\!\!\!/B\!\!\!\!/+\left(  \partial\!\!\!/B\!\!\!\!/\right)
\right]  ^{2}, \label{trace_2'}%
\end{equation}
so that we obtain the

\textbf{Theorem, nonabelian case}. \emph{Let }$D_{\mu}=\partial_{\mu}+B_{\mu}
$\emph{, where }$B_{\mu}$\emph{\ is a nonabelian vector field. Then:}
\begin{align}
\frac{1}{8}\mathrm{Tr}^{2}D\!\!\!\!/^{\,2}-\frac{1}{2}\mathrm{Tr}%
D\!\!\!\!/^{\,4}  &  =2(\partial\cdot B+B^{2})^{2}-\frac{1}{2}\mathrm{Tr}%
\left[  \left(  \partial\!\!\!/B\!\!\!\!/\right)
+B\!\!\!\!/B\!\!\!\!/\right]  ^{2}\nonumber\\
&  =\left(  (\partial_{\lbrack\mu}B_{\nu]})+[B_{\mu},B_{\nu}]\right)  ^{2},
\label{YMKE}%
\end{align}
\emph{where the traces are to be taken over the Dirac matrices.}

The kinetic energy of a nonabelian gauge field, using the spinorial covariant
derivative, can be written using the relation we have just derived. This part
of the lagrangian of a nonabelian gauge field is given, with the help of the
previous equation, by
\[
\frac{1}{2g^{2}}\widetilde{\mathrm{Tr}}\left(  \partial_{\lbrack\mu}A_{\nu
]}+[A_{\mu},A_{\nu}]\right)  ^{2}=\frac{1}{2g^{2}}\widetilde{\mathrm{Tr}%
}\left(  \frac{1}{8}\mathrm{Tr}^{2}D\!\!\!\!/^{\,2}-\frac{1}{2}\mathrm{Tr}%
D\!\!\!\!/^{\,4}\right)  .
\]

\subsection{Construction of a generalized Yang-Mills theory.}

The group-theoretical setup for a GYMT is the same as for a YMT. We begin with
a non-abelian local Lie group. Again for concreteness we choose the group to
be $SU(N).$ Just as we did when constructing the YMT we put the gauge fields
in the adjoint, so there must be $N$ of them. We put the fermions in the
fundamental $\mathbf{N}$, and use $\mathbf{N\times N}^{\ast}=\mathbf{Adj+1}$
to construct the invariant fermion energy term $\bar{\psi}iD\!\!\!\!/\psi.$

To construct the GYMT we generalize the transformation of the gauge field as
follows: to every generator in the Lie group we choose one gauge field that
can be either vector or scalar, so that if we choose $N_{V}$ generators to be
associated with an equal number of vector gauge fields and $N_{S}$ to be
associated with an equal number of scalar gauge fields then $N_{V}+N_{S}=N$.
If we choose a certain Lie group $G$ to base the GYMT on, then the maximal
subgroups of $G$ are the possible different GYMTs for that choice of group.
Both the form of the covariant derivative as well as the chiral structure of
the theory are determined by this choice.

Let say that we have decided upon a particular maximal subgroup $G_{M},$ so
that
\[
G\supset G_{M}.
\]
The way we construct the GYMT, if we restrict appropriately restrict the
transformation group, we will have a normal YMT with $G_{M}$ as gauge group.
The constructions proceeds as follows. We define the generalized covariant
derivative $D$ by taking each one of the generators and multiplying it by one
of its associated gauge fields and summing them together. The result is
\begin{equation}
D\equiv\partial\!\!\!/+A\!\!\!/+\Phi\label{der_na}%
\end{equation}
where
\begin{equation}%
\begin{array}
[c]{ll}%
A\!\!\!/=\gamma^{\mu}A_{\mu}=ig\gamma^{\mu}A_{\mu}^{a}T^{a}, & a=1,\ldots
,N_{V},\\
\Phi=\gamma^{5}\varphi=-g\gamma^{5}\varphi^{b}T^{b}, & b=N_{V}+1,\ldots,N,
\end{array}
\label{explicit}%
\end{equation}
and the generators $T^{a},$ $a=1,\ldots,N_{V},$ are the generators of the
maximal subgroups. The scalar fields $\varphi^{b}$ are associated only to
those generators $T^{b}$ that are generators of the original group $G,$ but
not of the maximal subgroup $G_{M}.$ (Observe the difference between $A_{\mu}
$ and $A_{\mu}^{a}$, and between $\varphi$ and $\varphi^{b}$.)

We now define the transformation for the gauge fields to be
\begin{equation}
A\!\!\!/+\Phi\rightarrow U(A\!\!\!/+\Phi)U^{-1}-(\partial\!\!\!/U)U^{-1},
\label{vec_4}%
\end{equation}
from which one can conclude that
\begin{equation}
D\rightarrow UDU^{-1} \label{transformation}%
\end{equation}
To construct the lagrangian we require that it should:

\begin{itemize}
\item  contain only fermion fields and covariant derivatives;

\item  possess both Lorentz and gauge invariance;

\item  have units of energy to the fourth power.
\end{itemize}

\noindent There are not many lagrangians compatible with these requirements.
There is a little freedom left in that one can still choose the irrep for the
fermions to lie in. A generic such lagrangian is%

\begin{equation}
\mathcal{L}_{GYM}=\bar{\psi}iD\psi+\frac{1}{2g^{2}}\widetilde{\mathrm{Tr}%
}\left(  \frac{1}{8}\mathrm{Tr}^{2}D^{\,2}-\frac{1}{2}\mathrm{Tr}%
D^{\,4}\right)  , \label{lag_na}%
\end{equation}
where the trace with the tilde is over the Lie group matrices and the one
without it is over matrices of the spinorial representation of the Lorentz
group. The additional factor of 1/2 that the traces of (\ref{lag_na}) have
with respect to (\ref{teo_1}) comes from the normalization given by
(\ref{normhalf}), the usual one in non-abelian YMTs.

The expansion of this lagrangian into component fields results in expressions
that are traditional in Yang-Mills theories. One has to substitute
(\ref{der_na}) in (\ref{lag_na}), and work out the algebra. It is convenient
to first get the intermediate result
\begin{align}
\frac{1}{16}\mathrm{Tr}^{2}D^{\,2}-\frac{1}{4}\mathrm{Tr}D^{\,4}  &  =\left(
(\partial\cdot A)+A^{2}\right)  ^{2}-\mathrm{Tr}\left(  (\partial
\!\!\!/A\!\!\!/)+A\!\!\!/A\!\!\!/\right)  ^{2}\nonumber\\
&  -\frac{1}{4}\mathrm{Tr}\left(  (\partial\!\!\!/\Phi)+\{A\!\!\!/,\Phi
\}\right)  ^{2}, \label{intermediate}%
\end{align}
where the curly brackets denote an anticommutator. Notice in this expression
that the differentiation operators are restrained, and that the $\gamma^{5}$
present in the scalar boson $\Phi$ has the dual essential function of ensuring
that the partials become restrained and the anticommutators commutators.

To finish the calculation we substitute (\ref{explicit}) in
(\ref{intermediate}) and in the fermionic term of (\ref{lag_na}) to obtain
\begin{align}
\mathcal{L}_{GYM}  &  =\bar{\psi}(i\partial\!\!\!/+A\!\!\!/)\psi-g\bar{\psi
}i\gamma^{5}\varphi^{b}T^{b}\psi+\frac{1}{2g^{2}}\widetilde{\mathrm{Tr}%
}\left(  \left(  \partial_{\lbrack\mu}A_{\nu]}+[A_{\mu},A_{\nu}]\right)
^{2}\right) \nonumber\\
&  +\frac{1}{g^{2}}\widetilde{\mathrm{Tr}}\left(  \left(  \partial_{\mu
}\varphi+i[A_{\mu},\varphi]\right)  ^{2}\right)  . \label{expanded}%
\end{align}
These are familiar structures: the first term on the right looks like the
usual matter term of a gauge theory, the second like a Yukawa term, the third
like the kinetic energy of vector bosons in a Yang-Mills theory and the fourth
like the gauge-invariant kinetic energy of scalar bosons in the non-abelian
adjoint representation. If we restrict the transformation group $G$ to the
maximal subgroup $G_{M}$ then those terms would not just ``look'' like a YMT,
they would be a YMT. It is also interesting to observe that, if in the last
term we set the vector bosons equal to zero, then this term simply becomes
$\sum_{b,\mu}\frac{1}{2}\partial_{\mu}\varphi^{b}\partial^{\mu}\varphi^{b}$,
the kinetic energy of the scalar bosons. We have constructed a generic
non-abelian gauge theory with gauge fields that can be either scalar or vector.

\subsection{Chiral structure.}

We have called the expanded lagrangian ``generic'', because it is the most
common form of a GYMT lagrangian; nevertheless, different choices of irrep for
the fermions will make its form vary slightly. Furthermore, we have not gone
into the chiral structure detail. Let us give an schematic example. Suppose
the maximal subgroup $G_{M}$ is composed of two subgroups. The spinor $\psi$
is then going to have some elements with right and some with left chiralities,
depending on which one of the subgroups of the maximal subgroup $G_{M}$ is
acting upon the element. If, for the example, the connection looks like this:
\[
\left(
\begin{array}
[c]{cc}%
\text{vector fields} & \text{scalar fields}\\
\text{scalar fields} & \text{vector fields}%
\end{array}
\right)  ,
\]
then the fermion interaction term in the lagrangian term will look like this:
\begin{equation}
\left(
\begin{array}
[c]{cc}%
\bar{\psi}_{L} & \bar{\psi}_{R}%
\end{array}
\right)  \left(
\begin{array}
[c]{cc}%
B\!\!\!\!/ & \phi\\
\phi^{\dagger} & C\!\!\!\!/
\end{array}
\right)  \left(
\begin{array}
[c]{c}%
\psi_{R}\\
\psi_{L}%
\end{array}
\right)  =\left(
\begin{array}
[c]{c}%
iB\!\!\!\!/\psi_{R}+i\phi\psi_{L}\\
-i\phi^{\dagger}\psi_{R}+iC\!\!\!\!/\psi_{L}%
\end{array}
\right)  .\label{structureb}%
\end{equation}
In this expression $\bar{\psi}_{L}\equiv\overline{\psi_{R}}$ and $\bar{\psi
}_{R}\equiv\overline{\psi_{L}}$, and we have written $B\!\!\!\!/$ and
$C\!\!\!\!/$ to generically represent the vector gauge fields of the two
subgroups that make up $G_{M},$ and $\phi$ to generically represent the scalar
gauge fields. Notice that in the expansion of this matrix product the terms
$\bar{\psi}_{L}B\!\!\!\!/\psi_{R}$ and $\bar{\psi}_{R}C\!\!\!\!/\psi_{L}$ have
their respective spinor fields with opposite chirality, and that the terms
$\bar{\psi}_{R}\phi\psi_{R}$ and $\bar{\psi}_{L}\phi\psi_{L}$ have their
respective spinor fields with the same chiralities, as is usual in Yukawa terms.

Terms like $\bar{\psi}_{L}B\!\!\!\!/\psi_{R}$ reflect the usual meaning of
connection: $\psi_{R}$ suffers the effect of the connection and then it is
multiplied the conjugate of its original self. On the other hand, a term like
$\bar{\psi}_{L}\phi\psi_{L}$ shows that the new part of the connection is more
like the space inversion discrete transformation in its effect than the usual
connection. It does not involve a vector index to relate to a path, and it
changes the chirality of the spinor, relating the left and right hand degrees
of freedom, like the mentioned transformation.

\section{A generalized Yang-Mills theory for Glashow-Weinberg-Salam: $U(3).$}

$SU(3)$ contains $U_{Y}(1)\times SU_{I}(2)$ as a maximal subgroup, so one
wonders if it could be a unification group for the GWSM. Its diagonal
generators correctly assign the hypercharge and isospin quantum numbers to the
electron and neutrino, but give the wrong hypercharge value to the Higgs
bosons because of the sign of the last component of the hypercharge generator.
Even worse, it does not make phenomenological sense at all to try to unify the
electroweak model using $SU(3),$because of the proliferation of physically
inexistent vector bosons. Yet some work has been done on $SU(3)$ as a
unification group$^{4},$ but our interest here is more along the lines of
Y.~Ne'eman and D.B.~Fairlie,$^{5,6}$ that considered using the graded
group$^{7}$ $SU(2/1)$, putting in the adjoint both the vector and the Higgs
bosons of the GWSM. This way there are no extra vector bosons arriving from
the unification; instead, the Higgs bosons nicely take those places.

The use of graded groups has, at least, two very serious problems. First, it
gives anticommuting properties to the Higgs bosons. Second, due to the
definition of the supertrace, at least one of the vector boson kinetic energy
terms has the wrong sign.

The simpler choice of the non-graded group $SU(3)$ was forgone on the basis
that it resulted in an incorrect prediction of the value of the Higgs boson's
hypercharge. However, it was later pointed out that $U(3)$ has a non-standard
representation that does give the correct hypercharge to the Higgs.$^{8}$ The
additional vector boson using $U(3)$ brings in, turns out to automatically
decouple from the rest of the model and, interestingly enough, all the terms
of the GWSM, except the Higgs boson potential $V(\varphi),$ come out correctly
from just two terms and two irreps, one for bosons and one for fermions. So
the door is opened to use scalar bosons as part of the covariant derivative.

\subsection{Why $SU(3)$ does not work.}

The GWSM is the product of two groups, each one with its own coupling
constant. Thus $U_{Y}(1)$ has $g^{\prime}$ and $SU_{I}(2)$ has $g.$ According
to the model, they must be related by
\[
g^{\prime}=g\tan\theta_{W},
\]
where $\theta_{W}$ is Weinberg's angle. It's experimental value is $\theta
_{W}\approx29%
{{}^o}%
.$ If instead we take $\theta_{W}=30%
{{}^o}%
$ exactly, then the two coupling constants must obey
\begin{equation}
g^{\prime}=g/\sqrt{3}. \label{couplings}%
\end{equation}

The generator for $SU(3)$ that we want to associate with hypercharge is
\begin{equation}
T^{8}=\frac{1}{2}\lambda^{8}=\frac{1}{2\sqrt{3}}\operatorname*{diag}%
(1,1,-2),\label{T8}%
\end{equation}
and with isospin is
\[
T^{3}=\frac{1}{2}\lambda^{3}=\frac{1}{2}\operatorname*{diag}(1,-1,0).
\]
These generators correspond to the vector bosons $B\!\!\!\!/$ and
$A\!\!\!/^{3}$, that in the GWSM would belong to $U_{Y}(1)$ and $SU_{I}(2),$
respectively. The electric charge of a particle should be given in terms of
these two generators by the relation
\begin{equation}
Q=T_{3}+\frac{1}{2}Y.\label{KN}%
\end{equation}
Here we have introduced the hypercharge operator $Y$. We have followed
tradition and placed a factor of $\frac{1}{2}$ multiplying the hypercharge.
Associating the boson fields $A\!\!\!/^{1}$, $A\!\!\!/^{2}$, $\varphi^{4}$,
$\varphi^{5}$, $\varphi^{6}$ and $\varphi^{7}$ to the generators $T^{1}$,
$T^{2}$, $T^{4}$, $T^{5}$, $T^{6}$ and $T^{7}$, respectively, and defining
$A\!\!\!/_{-}\equiv A\!\!\!/^{1}-iA\!\!\!/^{2}$, $\Phi_{1}\equiv\Phi^{4}%
-i\Phi^{5}$, $\Phi_{2}\equiv\Phi^{6}-i\Phi^{7}$, the covariant derivative of
the group $SU(3)$ then looks like this:
\begin{equation}
D_{SU(3)}=i\partial\!\!\!/+\frac{1}{2}\left(
\begin{array}
[c]{ccc}%
A\!\!\!/_{3}+B\!\!\!\!//\sqrt{3} & A\!\!\!/_{-} & \Phi_{1}\\
A\!\!\!/_{+} & -A\!\!\!/_{3}+B\!\!\!\!//\sqrt{3} & \Phi_{2}\\
\Phi_{1}^{\dagger} & \Phi_{2}^{\dagger} & -2B\!\!\!\!//\sqrt{3}%
\end{array}
\right)  ,\label{adjoint}%
\end{equation}
where the $A$ and $B$ fields both contain, as shown in (\ref{explicit}), the
same coupling constant. Notice, however, that since the $B$ field belongs in
the maximal subgroup $U(1),$ and this group is going to be identified with
$U_{Y}(1),$ it is possible to include the factor of $\frac{1}{\sqrt{3}}$ in a
redefinition of the coupling constant of the $U(1),$ precisely with form
(\ref{couplings}). Thus the $T^{8}$ generator is multiplied by $2\sqrt{3}$
leaving the appropriate operator for the hypercharge quantum number:
\begin{equation}
Y^{\prime}=2\sqrt{3}T^{8}=\operatorname*{diag}(1,1,-2).\label{Y'}%
\end{equation}
As we shall very soon see, not only $T^{8}$, but the other diagonal generators
that show in the GYMT for the GWSM and have a nonzero $(3,3)$ component
fortunately always have a normalization coefficient of $1/\sqrt{3}$ which will
allow us to use $g^{\prime}$ instead of $g$, and naturally mimic the two
coupling constant structure of the GWSM.

If we go ahead and make the identification of the $Y^{\prime}$ operator we
have constructed with the hypercharge operator, and of the $SU(3)$ $T^{3}$
generator with the isospin operator, then the electric charge operator is
given by
\begin{equation}
Q^{\prime}=\operatorname*{diag}(1,0,-1)\text{,} \label{Q'}%
\end{equation}
according to (\ref{KN}). The GWSM fermions must then go in the chiral triplet
\begin{equation}
\psi_{SU(3)}=\left(
\begin{array}
[c]{c}%
e_{R}^{c}\\
\nu_{R}^{c}\\
e_{R}%
\end{array}
\right)  , \label{chiral}%
\end{equation}
since this way they obtain the correct charge eigenvalues from (\ref{Q'}).

Now let us find the hypercharge of the Higgs, which is given by the
coefficients of the field itself after commutation with the hypercharge
generator.
\begin{equation}
\lbrack Y^{\prime},\Phi]=\left[  \left(
\begin{array}
[c]{ccc}%
1 & 0 & 0\\
0 & 1 & 0\\
0 & 0 & -2
\end{array}
\right)  ,\left(
\begin{array}
[c]{ccc}%
0 & 0 & \Phi_{1}\\
0 & 0 & \Phi_{2}\\
\Phi_{1}^{\dagger} & \Phi_{2}^{\dagger} & 0
\end{array}
\right)  \right]  =\left(
\begin{array}
[c]{ccc}%
0 & 0 & 3\Phi_{1}\\
0 & 0 & 3\Phi_{2}\\
-3\Phi_{1}^{\dagger} & -3\Phi_{2}^{\dagger} & 0
\end{array}
\right)  \label{skew}%
\end{equation}
So we get the value 3 and not 1, that we know from the GWSM is the correct
value. We conclude that it is not possible to get the GWSM from $SU(2/1).$

Alternatively, it is possible to derive this result directly from Dynkin
methods.$^{8}$ Thus for $SU(3)\supset U_{Y}(1)\times SU_{I}(2)$ one has that
$8=1_{0}+2_{3}+\bar{2}_{-3}+3_{0}$, where, for example, in $2_{3}$ the 2 is
the isospin multiplet and the subindex 3 the hypercharge. This means that the
$SU(3)$ theory has four vector bosons (the three $A^{a}$ and the $B)$ and four
scalar bosons (the two complex Higgs), and that the hypercharge of the Higgs
is 3, like we calculated above.

\subsection{Why $SU(2/1)$ does not work, either.}

So it becomes clear why graded algebras were invoked. The $SU(2/1)$ graded
algebra has eight generators, just like $SU(3)$, and all of them are precisely
the same as this Lie group's except for
\begin{equation}
T^{0}=\frac{1}{2\sqrt{3}}\operatorname*{diag}(1,1,2), \label{T0}%
\end{equation}
that differs from $T^{8}$ in the sign of the two. The point is that with this
generator one then defines a different hypercharge operator
\begin{equation}
Y=2\sqrt{3}T^{0}=\operatorname*{diag}(1,1,2) \label{Y}%
\end{equation}
that results in a the correct hypercharge assignment for the Higgs:
\begin{equation}
\lbrack Y,\Phi]=\left[  \left(
\begin{array}
[c]{ccc}%
1 & 0 & 0\\
0 & 1 & 0\\
0 & 0 & 2
\end{array}
\right)  ,\left(
\begin{array}
[c]{ccc}%
0 & 0 & \Phi_{1}\\
0 & 0 & \Phi_{2}\\
-\Phi_{1}^{\dagger} & -\Phi_{2}^{\dagger} & 0
\end{array}
\right)  \right]  . \label{skewskew}%
\end{equation}

In spite of this promising beginning, it turns out that there are two
fundamental problems about using graded groups. They are:

\begin{itemize}
\item  the trace of their tensors of $SU(2/1)$ is not group-invariant. Only
the supertrace is, but supertraces are not positive definite (they multiply by
$-1$ some of the diagonal elements), and thus at least one of the kinetic
energies of the vector bosons is going to have the wrong sign;

\item  the parameters that multiply the $T^{4},$ $T^{5},$ $T^{6}$ and $T^{7}$
generators of $SU(2/1)$ have to be Grassmann numbers, so that the Higgs fields
have to be anticommuting among themselves, in total disagreement with the
fundamentals of quantum field theory, that require scalar bosons to be commuting.
\end{itemize}

\noindent These problems are very serious, and with time interest in the model
waned.$^{9}$

\subsection{The difference $U(3)$ makes.}

While it is not possible to construct a GYMT for the GWSM using $SU(3),$ it is
possible to do it using $U(3).$ The generators of this group are the same as
those of $SU(3)$, but with an extra one that has to have a nonvanishing trace.
To see the reason for this let $U=\exp(i\omega^{a}T^{a})$, and notice that
$\det U=\exp(i\omega^{a}\operatorname*{Tr}T^{a})$. Since for $U(3)$ one has
that, in general, $\det U\neq+1$, then at least one of the $T^{a}$ has to be
traceless. Take this operator to be
\begin{equation}
T^{9}=\frac{1}{\sqrt{6}}\operatorname{diag}(1,1,1). \label{T9}%
\end{equation}
With the two generators $T^{9}$ and $T^{8}$ it is possible to institute a new
representation of $U(3).$ This new representation would contain the generators
$T^{1},$ ..., $T^{7}$ of $SU(3)$, the $SU(2/1)$ generator $T^{0}$ seen in
(\ref{T0}), and a new generator
\begin{equation}
T^{10}=\frac{1}{\sqrt{6}}\operatorname{diag}(1,1,-1). \label{T10}%
\end{equation}
The generators $T^{0}$ and $T^{10}$ can be rewritten as a linear combination
of the two original $U(3)$ generators $T^{8}$ and $T^{9},$ as follows:
\begin{align*}
T^{10}  &  =\frac{2\sqrt{2}}{3}T^{8}+\frac{1}{3}T^{9},\\
T^{0}  &  =-\frac{1}{3}T^{8}+\frac{2\sqrt{2}}{3}T^{9}.
\end{align*}
Notice that $T^{10}$ and $T^{0}$ are orthogonal under (\ref{normhalf}).

The nine generators must correspond to nine boson fields. The boson $\Upsilon$
we are to associate with the new generator $T^{10}$ should be scalar, since
this generator is not part of the maximal subgroup. What we are seeing in our
example is that the original group has a maximal subgroup $G_{M}=U(1)\times
U(1)\times SU(2)$ in the sense that it is both a subgroup and there is no
subgroup larger than it, that is still properly contained in $G.$ And what we
are forced to do to get the quantum numbers right is to take a linear
combination of those two $U(1)$'s, which is equivalent to taking the original
representation and using fields that are a mixture of scalar and vector
bosons, something like $A_{\mu}\gamma^{\mu}+\phi\gamma^{5}.$ It is interesting
that a similar thing happens again when we attempt grand unification$.$

From a phenomenological point of view it is clear we must take the new boson
$\Upsilon$ to be a scalar, since this way it is more likely it will decouple
from the other scalar bosons and the fermions. A vector boson will inevitably
couple with the other vector bosons and show up phenomenologically. This
decoupling of the scalar bosons from each other, and of the diagonal scalar
bosons from the fermions, is a remarkable property of GYMTs, and we shall
study it next for the case of the GWSM.

\subsection{Emergence of the Glashow-Weinberg-Salam model from $U(3).$}

Let us briefly review what has happened, but now using branching rules for
maximal subalgebras.$^{10}$ If one takes $SU(3)$ as GYMT then $SU(3)\supset
U_{Y}(1)\times SU_{I}(2).$ The fermions are then placed in the fundamental
\textbf{3} that has a branching rule $\mathbf{3=1}_{-2}\mathbf{+2}_{1}$ (the
hypercharge appears as subindex), that is, we end up with an isospin singlet
with $Y=-2$ and an isospin doublet with $Y=1.$ This implies the fermion
multiplet has to be chiral and of the form (\ref{chiral}). The problem in this
case lies elsewhere. The bosons are in the adjoint \textbf{8} that has the
branching rule $\mathbf{8=1}_{0}\mathbf{+2}_{3}\mathbf{+\bar{2}}%
_{-3}\mathbf{+3}_{0},$ so that we end up with a vector boson singlet, a vector
boson triplet, and a complex scalar doublet, the usual boson spectrum in the
GWSM. However, as we mentioned, this picture is unacceptable because the
hypercharge of the Higgs doublet is $Y=3$, as can be seen from the branching
rule. So, instead, we take the group $U(3)$ with the maximal subgroup
$U(3)\supset U_{Z^{\prime}}(1)\times U_{Y^{\prime}}(1)\times SU_{I}%
(2)\rightarrow U_{Z}(1)\times U_{Y}(1)\times SU_{I}(2),$ where we have called
$Z^{\prime}$ and $Y^{\prime}$ the quantum numbers associated with the gauge
bosons of the two $U(1)$'s, and $Z$ and $Y$ are generators that are linear
combinations of those numbers and give a particle spectrum with the same
quantum numbers of the GWSM. The electric charge operator that obtains through
the use of the hypercharge $Y$ of equation (\ref{Y}) is given by
\[
Q=\operatorname*{diag}(1,0,1)\text{,}%
\]
that implies that the fermion multiplet is the nonchiral
\begin{equation}
\psi_{U(3)}=\left(
\begin{array}
[c]{c}%
e_{R}^{c}\\
\nu_{R}^{c}\\
e_{L}^{c}%
\end{array}
\right)  . \label{nonchiral}%
\end{equation}
This choice of hypercharge assigns to all the particles that come out of the
$U(3)$ GYMT the correct quantum numbers of the GWSM.

Let us see how it is that the extra scalar boson $\Upsilon$ decouples in this
case. From an inspection of (\ref{expanded}) we see scalar bosons in GYMT do
not interact among themselves. Those terms seem to be included in
(\ref{lag_na}), but then disappear for algebraic reasons and are absent in
(\ref{expanded}). Scalar bosons only appear in the expanded generalized
lagrangian in the Yukawa term and in the gauge invariant kinetic energy term.
The interactive part of this term is proportional to%
\begin{equation}
\widetilde{\operatorname*{Tr}}\left(  \left(  [A_{\mu},\Phi]\right)
^{2}\right)  . \label{commutator}%
\end{equation}
The positioning of the fields in the adjoint representation is illustrated in
the matrix:
\[
\left(
\begin{array}
[c]{cc}%
\text{Vector fields and }\Upsilon & \text{Higgs fields }\varphi\\
\text{Higgs fields }\varphi & \text{Vector fields and }\Upsilon
\end{array}
\right)
\]
The commutator $[A_{\mu},\Phi]$ is certainly not zero when $\Phi$ is one of
the usual GWSM Higgs bosons $\varphi$ because these bosons occupy places off
the block diagonal, but when $\Phi$ is the scalar boson $\Upsilon,$ that only
has components along the diagonal and that is proportional to the identity
matrix in each of those boxes, then the commutator (\ref{commutator}) has to
be zero.

The decoupling of the $\Upsilon$ from the fermions is due to similar reasons.
The scalar bosons only couple to the fermions through the Yukawa terms, which
are proportional to $\bar{\psi}\Phi\psi.$ If $\Phi$ is one of the Higgs bosons
$\varphi$ then it occupies a block that is off the matrix' diagonal and so it
stands between fermion spinors of the same chirality. But the $\Upsilon$
connects spinors of the opposite chirality and thus this term is zero.

\subsection{The lagrangian of the Glashow-Weinberg-Salam model.}

The covariant derivative of the $U(3)$ GYMT we have been developing has the
form
\[
D_{SU(3)}=i\partial\!\!\!/+\frac{1}{2}\left(
\begin{array}
[c]{cc}%
igA\!\!\!/^{a}\sigma^{a}+g^{\prime}iB\!\!\!\!/ & -\sqrt{2}\gamma^{5}%
\hat{\varphi}\\
-\sqrt{2}\gamma^{5}\hat{\varphi}^{\dagger} & 2ig^{\prime}B\!\!\!\!/
\end{array}
\right)  ,
\]
where $(\sigma^{a})=(\sigma^{1},\sigma^{2},\sigma^{3})$ are the Pauli
matrices, and
\[
\hat{\varphi}=\left(
\begin{array}
[c]{c}%
\Phi_{1}\\
\Phi_{2}%
\end{array}
\right)  =\frac{1}{\sqrt{2}}\left(
\begin{array}
[c]{c}%
\varphi^{4}-i\varphi^{5}\\
\varphi^{6}-i\varphi^{7}%
\end{array}
\right)  .
\]
The GWSM lagrangian follows from the application of this covariant derivative
to lagrangian (\ref{expanded}):
\begin{align*}
\mathcal{L}_{GWSM}  &  =\overline{\theta_{R}^{c}}(i\partial\!\!\!/-\frac{1}%
{2}gA\!\!\!/^{a}\sigma^{a}\mathbf{-}\frac{1}{2}g^{\prime}B\!\!\!\!/)\theta
_{R}^{c}+\overline{e_{L}^{c}}(i\partial\!\!\!/-g^{\prime}B\!\!\!\!/)e_{L}%
^{c}\\
&  +i\frac{\sqrt{2}}{2}g\overline{e_{L}^{c}}\hat{\varphi}^{\dagger}\theta
_{R}^{c}-i\frac{\sqrt{2}}{2}g\overline{\theta_{R}^{c}}\hat{\varphi}e_{L}%
^{c}-\frac{1}{4}\mathbf{A}_{\mu\nu}\mathbf{\cdot A}^{\mu\nu}\mathbf{-}\frac
{1}{4}B_{\mu\nu}B^{\mu\nu}\\
&  +\left|  (\partial_{\mu}+\frac{1}{2}igA_{\mu}^{a}\sigma^{a}-\frac{1}%
{2}ig^{\prime}B_{\mu})\hat{\varphi}\right|  ^{2}+\frac{1}{2}(\partial_{\mu
}\Upsilon)^{2},
\end{align*}
where we have employed the fermion multiplet $\psi=$(\ref{nonchiral}) and we
have defined the Higgs doublet $\theta_{R}^{c}=\left(
\begin{array}
[c]{c}%
e_{R}^{c}\\
\nu_{R}^{c}%
\end{array}
\right)  .$ It is the usual GWSM lagrangian's, except that the Higgs potential
$V(\hat{\varphi})$ is missing. The extra scalar boson does not interact at all.

\section{Grand unification using generalized Yang-Mills theories.}

In this section we shall describe how to build a GUT using GYMTs, and
construct one such example based on the GYMT $SU(6),$ that is not that much
larger group than $SU(5)$. It would seem at first sight that the logical
choice for a generalized GUT would be to use the maximal subgroup
$SU(6)\supset SU(3)\times SU(3)\times U(1),$ because we know that a $SU_{C}(3)
$ is needed to model quantum chromodynamics.$^{11}$ However, this choice is a
step in the wrong direction and makes the algebra very messy. The correct way
is to take $SU(6)\supset SU(5)\times U(1),$ which basically results in the
usual $SU(5)$ GUT.$^{12}$ We shall not cover other GYMT grand unification
schemes here.

A word of clarification: in general there is a difference in the
``unification'' brought about in a GUT or a GYMT. In a GUT two or more forces
are unified choosing a large Lie group that has as products in one of its
maximal subgroups the Lie groups that represent the forces being unified.
Usually the breaking of the symmetry into the original forces is achieved
through a Higgs field multiplet with one component that has a nonzero VEV. In
the case of a GYMT the ``unification'' that is achieved includes the usual one
we have just described \emph{plus} an additional, rather technical one, that
consists in having only two terms and two irreps in the lagrangian. Compare
this with, for example, the $SU(5)$ GUT. It has two irreps for the fermions,
one for the vector bosons, and two for the Higgs fields, for a total of five
irreps. The terms in the lagrangian are: the kinetic energy of the vector
bosons, the kinetic energy of the two types of Higgs, two types of Yukawa
terms, the kinetic energies of the fermions, and potentials for the Higgs
fields, for a total of nine types of terms. On the other hand, there is one
problem with GYMTs: there are no potentials for the Higgs fields.

\subsection{ The Yang-Mills $SU(5)$ GUT.}

We are going to briefly review the $SU(5)$ GUT, particularly aspects that are
of relevance to our topic. It is a Yang-Mills theory based on $SU(5)$ with the
vector bosons placed in the adjoint \textbf{24}, and the fermions in two
irreps which are the conjugate fundamental \textbf{\={5}} and an antisymmetric
\textbf{10}. The Higgs bosons that break the symmetry in the GWSM occupy
another fundamental \textbf{5}, and the Higgs bosons that give very high
masses to the unseen vector bosons occupy another adjoint \textbf{24}.

The breaking of the $SU(5)$ symmetry is into the $SU_{C}(3)\times
SU_{I}(2)\times U_{Y}(1)$ maximal subgroup. The branching rule for the vector
bosons is
\begin{equation}
\mathbf{24=(1,1)}_{0}\mathbf{+(3,1)}_{0}\mathbf{+(2,3)}_{-5}\mathbf{+(2,\bar
{3})}_{5}\mathbf{+(1,8)}_{0},\label{24}%
\end{equation}
where we have placed the hypercharge as a subindex. The 12 vector bosons with
hypercharge $\pm5$ are not seen phenomenologically, so they are presumed to
couple with the Higgs of the \textbf{24}. The branching rule for the Higgs
bosons is the same as for the vector bosons, since they occupy the same irrep.
Of these \textbf{24} Higgs only one could have a nonzero VEV and that is the
$(1,1)_{0},$ because it has zero electric and color charges. Otherwise the
vacuum would also have these quantum numbers, and it does not. The generator
associated with this field is thus proportional to the matrix
\[
\operatorname*{diag}(2,2,2,-3,-3),
\]
according to the rule of obtaining the quantum numbers of fields in a matrix
by taking its commutation with the diagonal generators. We shall see how this
same Higgs shows up quite unexpectedly in the $SU(6)$ GYMT.

The group $SU(5)$ has a symmetric \textbf{15} irrep. Is it possible to place
the fermions in this single irrep instead of using the two irreps
\textbf{\={5}} and \textbf{10}? The answer, within a usual Yang-Mills theory,
is no, the reason being that the quantum numbers of the particles would force
the irrep to be nonchiral, and thus, unacceptable. However, in a GYMT, not
only is it \emph{possible} to place all the fermions in the \textbf{15}, it is
\emph{required} by the theory as we shall immediately see.

\subsection{The $SU(6)\supset SU(5)\times U(1)$ generalized grand unified theory.}

\noindent The low-dimensional irreps of $SU(6)$ are the fundamental
\textbf{5}, the antisymmetric \textbf{15}, the symmetric \textbf{21} and the
adjoint \textbf{35}. Since we have \textbf{15} fermions the only choice we
have is the \textbf{15}, but this implies a nonchiral irrep. In a normal YMT
this would have been a headache, since we would need the usual\textbf{
5}$^{\ast}$ and \textbf{10} structure. But the \textbf{15} fermion irrep is
precisely what we need in a GYMT, since it implies a nonchiral irrep the way
we need it precisely.

\subsection{The boson term.}

We use the maximal subgroup $SU(6)\supset SU(5)\times U(1).$ The branching
rule for the \textbf{35} is
\[
\mathbf{35=1}_{0}\mathbf{+5}_{6}\mathbf{+\bar{5}}_{-6}\mathbf{+24}_{0},
\]
where we have written the quantum number associated with the $U(1)$ as a
subindex. The $\mathbf{24}_{0}$ contains the 24 vector bosons of the $SU(5)$
GUT, 12 of which are the vector bosons of the GWSM and quantum chromodynamcs
and the other 12 the $X$ carriers of the grand unification force. The two
$\mathbf{5}$'s are the Higgs bosons that later branch into the GWSM Higgs
doublet; they are exactly the same irrep we just examined in the $SU(5)$ GUTS,
but now branching from the $SU(6)$'s \textbf{35}. Therefore we call them the
$\varphi,$ again. Finally the $\mathbf{1}_{0}$ is the scalar boson we shall
call $\Omega,$ which corresponds to the generator of a quantum number we shall
call ultracharge. It has the form
\begin{equation}
T_{6}^{\,U}\propto\operatorname*{diag}(1,1,1,1,1,-5). \label{ultracharge}%
\end{equation}
Here we have a big difference between Yang-Mills and generalized Yang-Mills
grand unification: in the former case the heavy (with a mass of the order of
the grand unification energy scale) Higgs are in a $\mathbf{24},$ in the
latter there is only one and in a singlet. The covariant derivative is then,
schematically,
\[
D_{SU(6)}=\partial\!\!\!/+\left(
\begin{array}
[c]{cc}%
\mathbf{24}+\Omega & \varphi\\
\varphi^{\dagger} & -5\Omega
\end{array}
\right)  .
\]
Notice that the color and electric charge generators, which all have a zero
$(6,6)$ component and are diagonal in the $5\times5$ block, commute with
$T^{U}$ so that $\Omega$ is completely neutral and can take a VEV. This is
consistent with the fact it is a $SU(5)$ singlet.

\subsection{The fermion term.}

The fermions are in the irrep \textbf{15} that is an antisymmetrization of the
Kronecker product of two fundamental irreps, that is, $\mathbf{15=6\times
6}|_{a}.$ Let us construct the invariant term that contains the fermions. It
is going to be slightly different from the generic case we considered when we
gave the construction of GYMTs. The fermion Hilbert space in this case is the
$6\times6$ \emph{matrix} $\psi$ formed by contracting the fermion fields with
the Clebsch-Gordan matrix relating $\mathbf{6\times6}|_{a}$ with the
\textbf{15}. This happens to be an antisymmetric $6\times6$ matrix that
maintains its antisymmetry under the transformation $\psi\rightarrow U\psi
U^{T},$ or, equivalently, $\bar{\psi}\rightarrow U^{\ast}\bar{\psi}U^{-1}.$
The covariant derivative transforms as $D\rightarrow UDU^{-1},$ so the
gauge-invariant term must then be
\begin{equation}
\widetilde{\operatorname*{Tr}}\left(  \bar{\psi}iD\psi\right)  .
\label{fermion term}%
\end{equation}
To prove the invariance of this term notice that under a group
transformation,
\[
\widetilde{\operatorname*{Tr}}\left(  \bar{\psi}iD\psi\right)  \rightarrow
\widetilde{\operatorname*{Tr}}U^{\ast}\bar{\psi}U^{-1}UDU^{-1}U\psi
U^{T}=\widetilde{\operatorname*{Tr}}\left(  \bar{\psi}iD\psi\right)  ,
\]
since all the $U$'s cancel.

The fermion Hilbert space vector under the $SU(6)\supset SU(5)\times U(1)$
maximal subalgebra, looks like the following $6\times6$ antisymmetric matrix,
with the $\mathbf{5}_{R}$ being a column vector with 5 components and the
\textbf{10} a $5\times5$ antisymmetric matrix:
\begin{equation}
\psi=\left(
\begin{array}
[c]{cc}%
\mathbf{10}_{L} & \mathbf{5}_{R}\\
\mathbf{-5}_{R}^{T} & \mathbf{0}%
\end{array}
\right)  . \label{15}%
\end{equation}
Again this is a nonchiral multiplet, as it always seems to be the case when
GYMTs are involved.

\subsection{A revision of the SU(6) model.}

Till now we have followed the recipe for constructing a GYMT for $SU(6)$ and
have found that the $SU(5)$ GUT seems to follows from it. Does it really?
First of all, although both models have the same dynamics, the GYMT version
has additional symmetries that can be useful as custodial symmetries and have
important physical consequences. Second, there is a problem we shall now study
and has to do with one difference between the usual GUT and the GYMT.

The Higgs bosons that supply the grand unification mass scale occupy a
\textbf{24} in the $SU(5)$ GUT. Under the maximal subgroup $SU(5)\supset
SU_{C}(3)\times SU_{I}(2)\times U_{Y}(1)$ the branching rule for this irrep is
(\ref{24}), where the hypercharge as a subindex. It is usually assumed in GUTs
that these scalar bosons are not observed because their masses are of the
order of their VEVs, a situation that would make them unobservable at present.
The component that has the nonzero VEV is the $(1,1)_{0}$ Higgs, that
corresponds to the generator
\begin{equation}
T_{6}^{Y}\propto\operatorname*{diag}(2,2,2,-3,-3,0),\label{hypercharge}%
\end{equation}
because it has no color or electric charge. We have given this generator the
name $T_{6}^{Y}$ because it evidently fulfills the same function in $SU(6)$ as
the hypercharge generator $T^{Y}\propto\operatorname*{diag}(2,2,2,3,3)$ did in
$SU(5),$ aside from the additional dimension. But this means that the
hypercharge $B_{\mu}$ vector boson should also be assigned it$.$ The is not
quite our previous result, since we have the $\Omega$ assigned to the
$\propto\operatorname*{diag}(1,1,1,1,1,-5)$ generator. This difficulty has a
solution reminiscent of the way the original problem of finding a GYMT for
GWSM was handled. In that case, as we have seen, the $U(3)$ group was used
instead of $SU(3),$ and the correct model was obtained through a nonstandard
representation of the group generators that would have been impossible with
simply $SU(3)$. To solve our problem we have to resort to a similar maneuver:
we must associate with the $SU(6)$ hypercharge generator $T_{6}^{Y}$ a field
$\Xi$ that is a mixture of the $B_{\mu}$ hypercharge vector field and of the
$\Omega$ Higgs. This field is given by
\[
\Xi=B\!\!\!\!/+\gamma^{5}\Omega.
\]
This change is enough to solve the problem since now the leptoquarks acquire
the large grand unification mass, while at the same time there is still a
vector boson that generates the hypercharge quantum number. We revert to the
usual $SU(5)$ GUT.

Another interesting point to notice is that due to peculiar algebraic
properties of GYMTs the two different types of Higgs bosons do not couple to
each other, as can be appreciated from (\ref{expanded}).

\subsection{Serendipitous cancellations.}

The coupling between fermions and vector bosons is given by the block
structure of the matrices. This structure is dictated by the maximal subgroup
that has been chosen and afterwards there is basically no more freedom of
choice for us. Yet the results are very interesting. Another point that
deserves attention is the way the formalism manages the correct coupling of
fermions and Higgs bosons. The light Higgs bosons have to couple to the
fermions in order to achieve mass generation, but the heavy Higgs cannot since
that would give the fermions a large mass. The interaction term is given by
(\ref{fermion term}) with (\ref{15}) replaced in it. A careful analysis of the
178 interaction terms that result from taking the trace of the product of
matrices shows that all the terms that should vanish do, and the ones that do
not, have the correct couplings. It is remarkable result, and one that
involves many serendipitous cancellations.

\section{Geometrical considerations.}

The curvature of a YMT can be found as we have already done for a Riemann
manifold. We take a small closed curve $C$ as we did in Section 3 and
calculate the parallel displacement of a vector caused by a connection
$A_{\mu}$ (instead of the Christoffel symbol). The result is the same as the
one we obtained in that section; that is, that the curvature is the commutator
of the covariant derivative. In other words, the curvature of a YMT is given
by $R_{a\mu\nu}^{b}=-[D_{\mu},D_{\nu}]_{ab}$ or, omitting the matrix
subindices and using the usual symbol $(F_{\mu\nu})_{ab}=-R_{a\mu\nu}^{b}$,
as
\[
F_{\mu\nu}=[D_{\mu},D_{\nu}],
\]
where $D_{\mu}$ is given by (\ref{nonabelian}). In the YMT lagrangian the
kinetic energy of the gauge bosons is proportional to the gauge trace of the
square of the curvature
\[
\mathcal{L}_{YMT}=\frac{1}{2g^{2}}\widetilde{\operatorname*{Tr}}F_{\mu\nu
}F^{\mu\nu}=\frac{1}{2g^{2}}\widetilde{\operatorname*{Tr}}\left(
(\partial_{\lbrack\mu}A_{\nu]}+[A_{\mu},A_{\nu}])^{2}\right)  .
\]

Our interest is to generalize this mathematics to a GYMT. We already know from
(\ref{YMKE}) how to write the Yang-Mills kinetic energy using spinors. The
natural generalization would be then to say that the kinetic energy related to
the curvature square of a GYMT is given by
\[
\mathcal{L}_{GYMT}=\frac{1}{2g^{2}}\widetilde{\mathrm{Tr}}\left(  \frac{1}%
{8}\mathrm{Tr}^{2}D^{\,2}-\frac{1}{2}\mathrm{Tr}D^{\,4}\right)  .
\]
So the question is, what is the curvature in a GYMT? What is the equivalent of
a Yang-Mills' $F_{\mu\nu}$ for a such a theory?

\subsection{The generalized curvature.}

The quantity that we want to call the generalized curvature should have the
properties we have come to regard as indispensable. It has to be gauge and
Lorentz invariant and it should not have unrestrained derivatives. Due to the
scarceness of mathematical constructs in GYMTs, it can only be made of
covariant derivatives. The best candidate for a generalized curvature is
\begin{equation}
F=\frac{1}{4}\operatorname*{Tr}D^{2}-D^{2}. \label{generalized curvature}%
\end{equation}
(Following our previous convention, the trace is over the Dirac matrices.) We
can expand this quantity using (\ref{der_na}). The square of the covariant
derivative is
\begin{equation}
D^{2}=\partial^{2}+A\!\!\!/^{2}+\varphi^{2}+(\partial\!\!\!/A\!\!\!/)+2A\cdot
\partial+(\partial\!\!\!/\Phi)+\{A\!\!\!/,\Phi\} \label{D2}%
\end{equation}
and%

\begin{equation}
\frac{1}{4}\operatorname*{Tr}D^{2}=\partial^{2}+A^{2}+\varphi^{2}%
+\partial\cdot A+2A\cdot\partial. \label{trace D2}%
\end{equation}
It is clear that for a nonabelian vector gauge field $A^{2}$ and
$A\!\!\!/^{2}$ are not equal, nor is $\{A\!\!\!/,\Phi\}$ zero. It is
straightforward to calculate that
\begin{equation}
F=\partial\cdot A-(\partial\!\!\!/A\!\!\!/)+A\cdot
A-A\!\!\!/A\!\!\!/-(\partial\!\!\!/\Phi)-\{A\!\!\!/,\Phi\}. \label{expanded F}%
\end{equation}
Thus $F$ is composed of restrained differential operators. Since $D$
transforms as:
\[
D\rightarrow UDU^{-1}\text{ or }D\rightarrow SDS^{-1},
\]
under the gauge or Lorentz group, respectively, it is obvious that the
curvature would also transform as
\[
F\rightarrow UFU^{-1}\text{ or }F\rightarrow SFS^{-1}.
\]

The Lorentz trace of the square of the curvature is easy to calculate:
\begin{align*}
\operatorname*{Tr}F^{2}  &  =\operatorname*{Tr}\left(  (\frac{1}%
{4}\operatorname*{Tr}D^{2})^{2}-\frac{1}{4}(\operatorname*{Tr}D^{2}%
)D^{2}-\frac{1}{4}D^{2}(\operatorname*{Tr}D^{2})+D^{4}\right) \\
&  =\operatorname*{Tr}D^{4}-\frac{1}{4}(\operatorname*{Tr}D^{2})^{2}.
\end{align*}
Thus the kinetic energy of a GYMT, written in terms of the curvature, is
\[
\mathcal{L}_{GYMT}=\frac{1}{2g^{2}}\widetilde{\mathrm{Tr}}\left(  \frac{1}%
{8}(\operatorname*{Tr}D^{2})^{2}-\frac{1}{2}\operatorname*{Tr}D^{4}\right)
=-\frac{1}{4g^{2}}\widetilde{\mathrm{Tr}}\operatorname*{Tr}\left(
F^{2}\right)  .
\]
So with GYMTs it goes as with YMTs: the kinetic energy term of the lagrangian
can be written in terms of the square of the curvature.

\subsection{ The meaning of the generalized curvature.}

We proceed to calculate the parallel transport around a small closed path. For
a YMT parallel transport with a connection $A_{\mu}$ around a small
parallelogram made up of two vectors $x^{\mu}$ and $y^{\mu}$ can be calculated
in a way similar to our calculation of Subsection 3.3 in a Riemann manifold.
Here we are dealing with flat spacetime, but the connection $A_{\mu}$ still
does rotate vectors when they move. The area two-form is $\Delta S^{\mu\nu
}\equiv\frac{1}{2}x^{[\mu}y^{\nu]},$ as before. The product of two vectors
$x^{\mu}$ and $y^{\mu}$ is defined to be
\[
x\cdot y\equiv\frac{1}{4}\operatorname*{Tr}x\!\!\!/y\!\!\!/=x^{\mu}\eta
_{\mu\nu}y^{\nu},
\]
where $\ \eta_{\mu\nu}$ is the metric of flat spacetime.

The generalized product of vectors times curvature in our formalism, following
the pattern of substituting inner products by traces over contracted vectors,
would logically be
\[
S\cdot F\equiv\frac{1}{4}\mathrm{Tr}Fx\!\!\!/y\!\!\!/.
\]
Substituting for $F$ from (\ref{expanded F}), and with the help of
(\ref{list}) and the trace theorems of (\ref{list}) we can derive
\begin{align*}
S\cdot F  &  =\frac{1}{4}\operatorname*{Tr}\left(  \partial\cdot
A-(\partial\!\!\!/A\!\!\!/)+A\cdot A-A\!\!\!/A\!\!\!/-(\partial\!\!\!/\Phi
)-\{A\!\!\!/,\Phi\}\right)  x\!\!\!/y\!\!\!/\\
&  =\left(  x\cdot(\partial A)\cdot y-y\cdot(\partial A)\cdot x+x\cdot AA\cdot
y-y\cdot AA\cdot x\right) \\
&  =x^{\mu}y^{\nu}(\partial_{\lbrack\mu}A_{\nu]}+A_{[\mu}A_{\nu]})=S^{\mu\nu
}F_{\mu\nu},
\end{align*}
which is the usual Yang-Mills result. This means that the straightforward
guess gives the correct generalization, or, at least, is a step in the right direction.

\subsection{The curvature in five dimensions.}

The curvature we obtained for a GYMT is thus just the same as if we were
dealing with a usual YMT. This is not a satisfying situation; information has
been lost because of the lone $\gamma^{5}$ in the traces that makes many of
them zero. The scalar fields turns out to be a useless bystander.

The origin of the problem lies in the lack of symmetry between the covariant
derivative, that has five types of gauge fields ($A_{\mu}$ and $\varphi)$,
while only four types of derivatives ($\partial_{\mu}$). This asymmetry
carries over to the vector $x\!\!\!/$ that has no $\gamma^{5}$ component. This
situation results in many cancellations. From a mathematical point of view the
esthetically pleasing thing to do is to add a new coordinate, which we call
$x^{5},$ and is to be interpreted as some sort of new dimension. The
parallelogram around which to perform the parallel transport is then in a
five-dimensional space, and the spinor form of the coordinates is:
\begin{equation}
\hat{x}\equiv x\!\!\!/+ix^{5}\gamma^{5}\quad\mathrm{and}\quad\hat{y}\equiv
y\!\!\!/+iy^{5}\gamma^{5}. \label{fifth dimension}%
\end{equation}

It is important that
\begin{equation}
\frac{1}{4}\operatorname*{Tr}\hat{x}\hat{y}=x\cdot y-x^{5}y^{5},
\label{metric distance}%
\end{equation}
a result consistent with our ideas that adding a scalar multiplied by a
$\gamma^{5}$ makes sense both physically and mathematically. Notice that the
last term in (\ref{metric distance}) has a negative sign, so that its sign is
spacelike, not timelike. In this respect it is pertinent that the term
$x^{5}\gamma^{5}$ is not an observable, being antihermitian. Let us explain
this comment by considering the electromagnetic interaction term\-
\[
\mathcal{L}_{EM}=\bar{\psi}A\!\!\!/_{EM}\psi.
\]
The quantity $\mathcal{L}_{EM}$ is real. To see this we take its hermitian
conjugate, and through the following elementary sequence arrive at itself
again: $\mathcal{L}_{EM}^{\ast}=\psi^{\dagger}A\!\!\!/_{EM}^{\dagger}%
\gamma^{0}\psi=\psi^{\dagger}\gamma^{0}A\!\!\!/_{EM}\psi=\mathcal{L}_{EM},$
since $\gamma^{0}\gamma^{\mu\dagger}=\gamma^{\mu}\gamma^{0}.$ If instead of
$A\!\!\!/$ we had used $x\!\!\!/$ the result would have exactly the same.
However, if we had simply used $x^{5}\gamma^{5}$, without the $i,$ then
$(\bar{\psi}x^{5}\gamma^{5}\psi)^{\ast}=(\bar{\psi}x^{5}\gamma^{5}%
\psi)^{\dagger}=-\bar{\psi}x^{5}\gamma^{5}\psi,$ since $\gamma^{5\dagger
}=\gamma^{5}$ and $\gamma^{0}\gamma^{5}=-\gamma^{5}\gamma^{0}.$ This is the
reason that the last terms on the right in (\ref{fifth dimension}) require an
$i,$ with the implication that the fifth dimension is spacelike, not timelike.

When we perform parallel transport with the fifth dimension added the
situation changes drastically. What we have now is a five-dimensional
parallelogram, so that, calling $S_{5}$ the five-dimensional area element, the
change in vector due to parallel transport around $S_{5}$ is:
\[
S_{5}\cdot F=\frac{1}{4}\mathrm{Tr}F\hat{x}\hat{y}.
\]
With respect to the previous calculation with the four dimensional
parallelogram, this one has an extra term that we can calculate:
\begin{align*}
(S_{5}-S_{4})\cdot F  &  =\frac{1}{4}\operatorname*{Tr}F(x\!\!\!/y^{5}%
\gamma^{5}+x^{5}\gamma^{5}y\!\!\!/+x^{5}y^{5})\\
&  =-\frac{1}{4}\operatorname*{Tr}\left(  (\partial\!\!\!/\Phi)x\!\!\!/y^{5}%
\gamma^{5}+(\partial\!\!\!/\Phi)x^{5}\gamma^{5}y\!\!\!/+\{A\!\!\!/,\Phi
\}x\!\!\!/y^{5}\gamma^{5}+\{A\!\!\!/,\Phi\}x^{5}\gamma^{5}y\!\!\!/\right) \\
&  =\frac{1}{4}\operatorname*{Tr}\left(  (\partial\!\!\!/\varphi
)x\!\!\!/y^{5}-(\partial\!\!\!/\varphi)x^{5}y\!\!\!/+[A\!\!\!/,\varphi
]x\!\!\!/y^{5}-[A\!\!\!/,\varphi]x^{5}y\!\!\!/\right) \\
&  =x\cdot(\partial\varphi)y^{5}-y\cdot(\partial\varphi)x^{5}+x\cdot\lbrack
A,\varphi]y^{5}-y\cdot\lbrack A,\varphi]x^{5}\\
&  =(x^{\mu}y^{5}-x^{5}y^{\mu})\left(  (\partial_{\mu}\varphi)+[A,\varphi
]\right)  .
\end{align*}
This result has to be added to the previous one of $x^{\mu}x^{\nu}F_{\mu\nu},
$ and it is precisely This extra term employs the scalar field much in the
same way as the vector fields the vector fields of a YMT.

\section{Other implications of the new dimension.}

The use of five coordinates in $\hat{x}\equiv x\!\!\!/+x^{5}\gamma^{5}$ has an
immediate implication we must carefully consider now. If there is another
dimension there can be motion in that direction and so we must modify also the
covariant derivative. We have been using $D=\partial\!\!\!/+A\!\!\!/+\Phi,$
but instead now we are going to use
\begin{equation}
\hat{D}=D+i\gamma^{5}\partial_{5}=\partial\!\!\!/+i\gamma^{5}\partial
_{5}+A\!\!\!/+\Phi, \label{new derivative}%
\end{equation}
where $\partial_{5}=\partial/\partial x^{5}.$ We have preferred to give the
new coordinate the name $x^{5}$ and not $x^{4},$ because it has to be
associated with the $\gamma^{5}.$ This form of the covariant derivative is
more symmetrical, since both the partials and the gauge fields benefit from a
scalar term that is then multiplied by $\gamma^{5}.$ This new covariant
derivative changes some of our previous results.

\subsection{The new five-dimensional curvature.}

The curvature itself is affected by the new form the covariant derivative has.
Surprisingly, it differs by only one term from the previous four-dimensional
GYMT curvature. All the other terms either cancel or vanish according to the
formulae of (\ref{list}). The new curvature $\hat{F}$ is given by
\[
\hat{F}=\frac{1}{4}\operatorname*{Tr}\hat{D}^{2}-\hat{D}^{2},
\]
and
\begin{equation}
\hat{D}^{2}=D^{2}+i\gamma^{5}(\partial_{5}A\!\!\!/)+i\{\varphi,\partial
_{5}\}-\partial_{5}{}^{2}. \label{derivative square}%
\end{equation}
(Notice that the terms $\{\varphi,\partial_{5}\}+\partial_{5}{}^{2}$ are
Lorentz scalars. The reader that has followed closely previous typical
derivations may have observed that squared scalars of the covariant derivative
usually cancel in GYMT's calculations, so that probably there are not going to
be terms like $\partial_{5}\varphi$ in the result.) Taking the trace of the
previous equation we get
\begin{equation}
\frac{1}{4}\operatorname*{Tr}\hat{D}^{2}=\frac{1}{4}\operatorname*{Tr}%
D^{2}+i\{\varphi,\partial_{5}\}-\partial_{5}{}^{2}, \label{trace derivative}%
\end{equation}
so that
\begin{align}
\hat{F}  &  =F-i\gamma^{5}(\partial_{5}A\!\!\!/)\nonumber\\
&  =\partial\cdot A-(\partial\!\!\!/A\!\!\!/)+A\cdot
A-A\!\!\!/A\!\!\!/-(\partial\!\!\!/\Phi)-\{A\!\!\!/,\Phi\}-i\gamma
^{5}(\partial_{5}A\!\!\!/). \label{new curvature}%
\end{align}

We can now recalculate the parallel transport over a small five-dimensional
parallelogram. The result come out to be:
\begin{align*}
S_{5}\cdot\hat{F}  &  \equiv\frac{1}{4}\mathrm{Tr}\hat{F}\hat{x}\hat{y}\\
&  =\frac{1}{4}\mathrm{Tr}F\hat{x}\hat{y}+\frac{1}{4}%
\mathrm{\operatorname*{Tr}}i\gamma^{5}(\partial_{5}A\!\!\!/)(x\!\!\!/y^{5}%
\gamma^{5}+x^{5}\gamma^{5}y\!\!\!/)\\
&  =S_{5}\cdot F-i(\partial_{5}A_{\mu})(x^{\mu}y^{5}-x^{5}y^{\mu}).
\end{align*}
Putting all the terms together, the parallel transport is
\begin{equation}
S_{5}\cdot\hat{F}=S_{\mu\nu}(\partial_{\lbrack\mu}A_{\nu]}+A_{[\mu}A_{\nu
]})+S_{\mu5}\left(  (\partial_{\mu}\varphi)-i(\partial_{5}A_{\mu}%
)+[A,\varphi]\right)  \label{quintet}%
\end{equation}
where $S_{\mu5}=(x^{\mu}y^{5}-x^{5}y^{\mu}).$ This looks very much like the
curvature of a YMT in five dimensions, with the scalar fields $\varphi$ taking
the role of the fifth component of the gauge field $A_{\mu}$. The difference
is that in a real YMT in five dimensions each generator $T^{a}$ is multiplied
with all five $\mu=0,1,2,3,5$ components of the gauge field $A_{\mu}^{a},$
while in a GYMT a generator $T^{a}$ is multiplied with either the first four
components $\mu=0,1,2,3$ of the vector field or with an individual scalar
field. In a five-dimensional YMT the number of generators $T^{a}$ is the same
as the number of five-dimensional vector fields, while in a GYMT roughly half
the number of generators of the Lie group are associated with the
four-dimensional vector fields $A_{\mu}^{a},$ and half with the scalar fields
$\varphi^{b}$.

The interest of result (\ref{quintet}) is that it uses the extra scalar field.
This is evidence in favor of the five-dimensional quantity $\hat{x}.$ In the
next subsections the theme is further examined.

\subsection{A unexpected meaningful quantity.}

If one looks at the way the results of the previous subsection were obtained,
one realizes that GYMTs may possess an additional physical quantity that YMTs
do not. For GYMTs it makes sense to calculate a different kind of
``curvature'' that involves only one dimension instead of two.

Consider the new quantity $\mathrm{\operatorname*{Tr}}\hat{F}\hat{x},$ with
only one coordinate segment$.$ Using (\ref{new curvature}), (\ref{fifth
dimension}) and (\ref{list}) it is easy to verify that
$\mathrm{\operatorname*{Tr}}\hat{F}\hat{x}=0.$ However, it does make sense to
calculate the quantity $\mathrm{\operatorname*{Tr}}\hat{F}\hat{x}i\gamma^{5},$
that has an extra chirality matrix factor. If we were dealing with a YMT, then
$F$ would have only the terms $\partial\cdot A-(\partial
\!\!\!/A\!\!\!/)+A\cdot A-A\!\!\!/A\!\!\!/$, and $\mathrm{Tr}%
Fx\!\!\!/=\mathrm{Tr}Fx\!\!\!/i\gamma^{5}=0$. So within a YMT it does not make
sense to define quantities of this kind, but within a GYMT it does. Let us
calculate this new quantity:
\begin{equation}
\frac{1}{4}\mathrm{Tr}\hat{F}i\hat{x}\gamma^{5}=ix^{\mu}\left(  (\partial
_{\mu}\varphi)-i(\partial_{5}A_{\mu})+[A_{\mu},\varphi]\right)  .
\label{new quantity}%
\end{equation}
As it frequently happens in GYMTs, a formula that involves many terms ends up
with relatively few. Notice this quantity is Lorentz- and gauge-invariant, and
has no unrestrained coordinates. Every terms on the right includes a structure
that is present in GYMTs but is not in YMTs, so there can be no counterpart to
this quantity in the latter theories. It looks like a local gauge-invariant dilation.

\subsection{Changes in the lagrangian due to the fifth dimension.}

The extra term in the covariant derivative makes absolutely no difference in
the fermion sector. To verify this statement recall that this sector is
composed of the single term
\begin{equation}
\mathcal{L}_{F}=\bar{\psi}i\hat{D}\psi=\bar{\psi}iD\psi-\bar{\psi}\gamma
^{5}\partial_{5}\psi. \label{F5D}%
\end{equation}
The first term on the right is the usual one in GYMTs, and the second one
gives no contribution since
\[
\bar{\psi}\gamma^{5}\partial_{5}\psi=\bar{\psi}_{L}\gamma^{5}\partial_{5}%
\psi_{R}+\bar{\psi}_{R}\gamma^{5}\partial_{5}\psi_{L}=0,
\]
because of the chirality projectors$.$ This cancellation mechanism, that
impedes any contribution from $i\partial_{5}\gamma^{5}$ to the fermion sector,
is exactly the same one that annuls the interaction terms between the heavy
Higgs and the fermions.

To study the effect of the new covariant derivative in the bosonic sector
\begin{equation}
\mathcal{L}_{B}=\frac{1}{2g^{2}}\widetilde{\mathrm{Tr}}\left(  \frac{1}%
{8}\mathrm{Tr}^{2}\hat{D}^{\,2}-\frac{1}{2}\mathrm{Tr}\hat{D}^{\,4}\right)  ,
\label{B5D}%
\end{equation}
we first calculate the explicit form of $\hat{D}^{2}$:
\begin{align*}
\hat{D}^{2}  &  =\partial^{2}+\varphi^{2}+2A\cdot\partial+i\{\varphi
,\partial_{5}\}-\partial_{5}{}^{2}\\
&  +A\!\!\!/^{2}+(\partial\!\!\!A\!\!\!/)+(\partial\!\!\!/\Phi
)+\{A\!\!\!/,\Phi\}+i(\partial_{5}A\!\!\!/).
\end{align*}
From here we can immediately conclude taking the trace that
\[
\frac{1}{4}\operatorname*{Tr}\hat{D}^{2}=\partial^{2}+\varphi^{2}%
+2A\cdot\partial+i\{\varphi,\partial_{5}\}-\partial_{5}{}^{2}+A^{2}%
+\partial\!\!\!\cdot A.
\]
There are several terms involved in each of the two trace expressions between
parenthesis in (\ref{B5D}), but a scrutiny of the possible types of
combinations of products of $\gamma^{\mu}$'s and $\gamma^{5}$'s shows that
there is going to be much cancellation of terms between those two trace
expressions. With this in mind let us look at the following substraction:
\begin{align*}
\frac{1}{4^{2}}\operatorname*{Tr}\nolimits^{2}\hat{D}^{2}-\frac{1}{4}\hat
{D}^{4}  &  =\left(  (\partial\cdot A)+A^{2}\right)  ^{2}-\frac{1}%
{4}\operatorname*{Tr}\left(  (\partial
\!\!\!/A\!\!\!/)+A\!\!\!/A\!\!\!/\right)  ^{2}\\
&  -\frac{1}{4}\operatorname*{Tr}((\partial\!\!\!/\Phi)(\partial
\!\!\!/\Phi)+(\partial\!\!\!/\Phi)\{A\!\!\!/,\Phi\}+(\partial\!\!\!/\Phi
)(i\gamma^{5}\partial_{5}A\!\!\!/)\\
&  +\{A\!\!\!/,\Phi\}(\partial\!\!\!/\Phi)+\{A\!\!\!/,\Phi\}\{A\!\!\!/,\Phi
\}+\{A\!\!\!/,\Phi\}(i\gamma^{5}\partial_{5}A\!\!\!/)\\
&  +(i\gamma^{5}\partial_{5}A\!\!\!/)(\partial\!\!\!/\Phi)+(i\gamma
^{5}\partial_{5}A\!\!\!/)\{A\!\!\!/,\Phi\}+(i\gamma^{5}\partial_{5}%
A\!\!\!/)(i\gamma^{5}\partial_{5}A\!\!\!/))\\
&  =\frac{1}{2}\left(  (\partial_{\lbrack\mu}A_{\nu]})+[A_{\mu},A_{\nu
}]\right)  ^{2}+\left(  (\partial_{\mu}\varphi)-i(\partial_{5}A_{\mu}%
)+[A^{\mu},\varphi]\right)  ^{2}.
\end{align*}
So we finally find the explicit form of the boson kinetic energy:
\begin{equation}
\mathcal{L}_{B}=\frac{1}{2g^{2}}\widetilde{\mathrm{Tr}}\left(  \left(
\partial_{\lbrack\mu}A_{\nu]}+[A_{\mu},A_{\nu}]\right)  ^{2}\right)  +\frac
{1}{g^{2}}\widetilde{\mathrm{Tr}}\left(  \left(  \partial_{\mu}\varphi
-i\partial_{5}A_{\mu}+[A_{\mu},\varphi]\right)  ^{2}\right)
\label{boson term}%
\end{equation}
We see that the only change in the bosonic sector due to the extra dimension
has been the addition of a new negative term in the kinetic energy of the
scalar bosons, the term is $-i(\partial_{5}A_{\mu})$.

\subsection{Five-dimensional formulation of the lagrangian.}

With the help of definitions
\begin{equation}
A_{5}\equiv-i\varphi,\qquad\hat{\gamma}^{5}\equiv i\gamma^{5},\qquad
\hat{\gamma}^{\mu}\equiv\gamma^{\mu}, \label{scalars in 5}%
\end{equation}
the covariant derivative in five dimensions can be written in the interesting
form
\begin{align*}
\hat{D}  &  =\partial\!\!\!/+i\gamma^{5}\partial_{5}+A\!\!\!/+\gamma
^{5}\varphi\\
&  =\partial\!\!\!/+\hat{\gamma}^{5}\partial_{5}+A\!\!\!/+\hat{\gamma}%
^{5}A_{5}\\
&  =\hat{\gamma}^{m}(\partial_{m}+A_{m}),\qquad m=0,1,2,3,5,
\end{align*}
where, as usual, a repeated index implies sum over its values. The expression
that appears in the boson term of a GYMT in five dimensions can be similarly
modified:
\[
(\partial_{\mu}\varphi)-i(\partial_{5}A_{\mu})+[A_{\mu},\varphi]^{2}=i\left(
\partial_{\lbrack\mu}A_{5]})+i[A_{\mu},A_{5}]\right)  ,
\]
so that (\ref{boson term}) takes the brief and convenient appearance
\[
\mathcal{L}_{B}=\frac{1}{2g^{2}}\widetilde{\mathrm{Tr}}\left(  \left(
\partial_{\lbrack m}A_{n]}+[A_{m},A_{n}]\right)  \left(  \partial^{\lbrack
m}A^{n]}+[A^{m},A^{n}]\right)  \right)  ,\qquad m,n=0,1,2,3,5,
\]
where we have used $A^{5}=-A_{5}$ and $\partial_{5}=-\partial^{5},$ as we
should if we are to be consistent with (\ref{metric distance}). Thus the boson
term of a GYMT with an extra dimension looks superficially like a YMT in five
dimensions. If we ignore completely the gauge structure of the theory, it
would be a YMT, at least in the boson sector. Besides this difference there is
another between these two types of theories. It hinges upon the fact that in a
YMT in a higher dimension it is necessary to consider the spinorial
representations available to that dimension, while for the variation of a GYMT
we have been working on there is no need for that, since the extra Dirac
matrix $\hat{\gamma}^{5}$ we have been using is just an antihermitian form of
the chirality matrix $\gamma^{5}$ of the original Dirac four-dimensional
matrix algebra.

The five-dimensional covariant derivative (\ref{new derivative}) can also be
written, using the new symbol $\hat{D}_{m}=\partial_{m}+A_{m},$ $m=0,1,2,3,5,
$ in the familiar-looking form
\[
\hat{D}=\hat{\gamma}^{m}\hat{D}_{m}.
\]
The gauge transformation is modified into the five-dimensional form
\begin{equation}
\hat{\gamma}^{m}A_{m}\rightarrow U\hat{\gamma}^{m}A_{m}U^{-1}-(\hat{\gamma
}^{m}\partial_{m}U)U^{-1}, \label{transformation_5}%
\end{equation}
or
\[
A_{m}\rightarrow UA_{m}U^{-1}-(\partial_{m}U)U^{-1}.
\]
Furthermore, using the new symbol $\hat{D}_{m}$ the boson term can be written
simply as
\[
\mathcal{L}_{B}=\frac{1}{2g^{2}}\widetilde{\mathrm{Tr}}\left(  ([\hat{D}%
_{m},\hat{D}_{n}])^{2}\right)  .
\]

\subsection{Effect of the extra partial derivative $\partial_{5}.$}

To find what is the effect of the extra partial derivative we must find the
equation of motion of the gauge fields. To this end we take the first-order
variation of the bosonic term of the lagrangian with respect to $\delta A_{m}
$ (which includes now the scalar fields with the label $A_{5}$), with the
result
\begin{align}
\delta\mathcal{L}_{B}  &  =\frac{1}{g^{2}}\widetilde{\mathrm{Tr}}\left(
F^{mn}\left(  \frac{\partial(\partial_{\lbrack m}A_{n]})}{\partial A_{p}%
}+\frac{\partial\lbrack A_{m},A_{n}]}{\partial A_{p}}\right)  \delta
A_{p}\right) \nonumber\\
&  =\frac{2}{g^{2}}\widetilde{\mathrm{Tr}}\left(  F^{mn}(-\overleftarrow
{\partial}_{m}\delta_{n}^{p}\delta A_{p}+\delta A_{p}\delta_{m}^{p}A_{n}%
+A_{m}\delta_{n}^{p}\delta A_{p})\right) \nonumber\\
&  =-\frac{2}{g^{2}}\widetilde{\mathrm{Tr}}\left(  (\partial_{m}F^{mp}%
-A_{n}F^{pn}-F^{mp}A_{m})\delta A_{p}\right) \nonumber\\
&  =-\frac{2}{g^{2}}\widetilde{\mathrm{Tr}}\left(  (\partial_{m}F^{mn}%
+A_{m}F^{mn}-F^{mn}A_{m})\delta A_{n}\right) \nonumber\\
&  =-\frac{2}{g^{2}}\widetilde{\mathrm{Tr}}(\partial_{m}F^{mn}+[A_{m}%
,F^{mn}])\delta A_{n}=0, \label{change}%
\end{align}
where
\begin{equation}
F_{mn}\equiv\partial_{\lbrack m}A_{n]}+[A_{m},A_{n}]=[\hat{D}_{m},\hat{D}%
_{n}]. \label{equation}%
\end{equation}
The equations of motion are then%

\[
\partial_{m}F^{mn}+[A_{m},F^{mn}]=0.
\]
Again, these equations look like YMT equations of motion, but they are not,
due to the differing gauge structure. Of course, there is the additional
difference of having $m=5$ be a compactified dimension.

\subsection{Ideas about the five-dimensional equations of motion.}

The nonlinear character of equations (\ref{equation}) make their solution
difficult. Furthermore other technical problems are involved in this problem,
due to the compactification of one of the dimensions. The natural hope is that
the least energy solution of these equations (or one of the least energy
solutions), can result in a VEV of the order of the mass of the grand
unification leptoquarks, say $10^{15}%
\operatorname{GeV}%
,$ for the scalar field$.\Omega.$ This field is hypothetically trapped in a
small circle with coordinate $x^{5}.$ Since we know that $1\hbar c\approx200%
\operatorname{MeV}%
$-fm, it is possible to conclude that the radius of the circle formed by the
compactified dimension is of the order of $\sim2\times10^{-29}%
\operatorname{cm}%
.$ Obviously we cannot notice motion on such a small scale. We shall not dwell
further in this topic. It goes without saying that it is crucial that this
last dimension does not change the correct phenomenology of the
four-dimensional GYMT, otherwise it would be necessary to do without it. But
it would be a shame, because the extra dimension adds both sense and balance
to the mathematics involved. For example, the extra dimension gives a clear
geometric meaning to the scalar fields, as was discussed in Section 8. It also
gives a clearer meaning to the gauge transformation of the boson fields,
equations (\ref{transformation}) and (\ref{transformation_5}).

\section{Final comments and criticisms.}

We have gone through some very suggestive information concerning the structure
of GYMTs. It would seem that these theories have relevance to particle
physics, judging from the way they allow esthetic presentations of the GWSM
and GUTs. However, both their mathematical and physical situations require
clarification. The mathematical structures involved are complicated, and
several points are not quite clear. We have gone in some detail into the
possibility of using an extra dimension in their presentation with very
interesting results. However, the correct way to proceed is still uncertain.
Possibly the extra dimension that is introduced into this kind of GYMT is
compactified, supplying a large mass scale.

From a physical point of view, one complaint is that these theories do not
present an explicit potential $V(\phi)$ that would explain the nonzero VEVs of
the Higgs fields $\phi$. In this they differ from the GWSM. On the other hand,
this potential is pretty much \emph{ad hoc} in that model. Another important
point is that the physical implications of the additional gauge symmetry that
GYMTs contain ($SU(6)$ for the GUT we studied, for example) have not been yet
properly evaluated.

In five dimensions GYMTs look like a YMT, but this is just a superficial
similarity, since the gauge structure is different. In this type of GYMT each
scalar field is taken to be the fifth component of a vector, and it always has
a generator of the Lie group assigned to it that is not the same one assigned
to the other components of the generalized gauge field. On the other hand, in
a YMT each component of a particular vector field is associated with the same
generator of the Lie group.

An interesting feature of GYMTs is that, even if branching rules tell us that
two scalar fields have the correct quantum numbers to couple, they are still
prevented to do so by the mathematics.

\bigskip

\noindent\textbf{References}

\begin{enumerate}
\item  H. Weyl, \emph{Z. Physik} \textbf{56}, 330 (1929).

\item  C.N. Yang and R.L. Mills, Phys. Rev. \textbf{96}, 96 (1954).

\item  S. Weinberg, \emph{Gravitation and Cosmology: Principles and
Applications of the General Theory of Relativity,} (John Wiley \& Sons, Inc.,
N. Y. 1973) p. 133.

\item  N. Atsuji, I. It\"{o}, S.Y. Tsai, T. Kimura and K. Furuya, \emph{Prog.
Theo. Phys.} \textbf{67,} 1149 (1982); M. Singer, R. Parthasarathy and
K.S.Viswanathan, \emph{Phys. Lett. }\textbf{125,} 63 (1983); F. Pisano and V.
Pleitez, \emph{Phys. Rev. }\textbf{D46, 410 }(1992); D. Ng, \emph{Phys. Rev.
}\textbf{D49, }4805\textbf{\ }(1994).

\item  Y. Ne'eman, \emph{Phys. Lett. }\textbf{B81, 190 }(1979); D.B. Fairlie,
\emph{Phys. Lett.}\textbf{\ B82,} 97 (1979).

\item  R. E. Ecclestone,\emph{\ J. Phys}. \textbf{A13,} 1395 (1980); R. E.
Ecclestone, \emph{Phys. Lett.} \textbf{B116,} 21 (1982).

\item  B. DeWitt, \emph{Supermanifolds}, Second ed., (Cambridge University
Press 1992, Cambridge N. Y).

\item  M. Chaves and H. Morales, \emph{Mod. Phys. Lett.} A\textbf{13,} 2021
(1998); M. Chaves, ``Some Mathematical Considerations about Generalized
Yang-Mills Theories'', in \emph{Photon and Poincar\'{e} Group}, ed. V. V.
Dvoeglazov, (Nova Science Publishers, New York, 1998) p. 326.

\item  P.H. Dondi and P.D. Jarvis, \emph{Phys. Lett.} B\textbf{84,} 75 (1979);
E.J. Squires, \emph{Phys. Lett.} B\textbf{82,} 395 (1979); J.G. Taylor,
\emph{Phys. Lett.} B\textbf{83}, 331 (1979); J.G. Taylor, \emph{Phys. Lett.}
B\textbf{84}, 79 (1979); J.G. Taylor, \emph{Phys. Rev. Lett.} \textbf{43,} 824
(1979); J.~Thierry-Mieg and Y.~Ne'eman, \emph{Il Nuovo Cimento} A\textbf{71,}
104 (1982); Y.~Ne'eman, \emph{Phys. Lett.} B\textbf{181,} 308 (1986); S. Naka,
H. Chujyou and H. Kan, \emph{Prog. Theo. Phys.} \textbf{85}, 643 (1991); H.
Kan and S. Naka, \emph{Prog. Theo. Phys}. \textbf{90,} 1161 (1993).

\item  R. Slansky, \emph{Phys. Rep.} \textbf{79}, 1 (1981), especially Table 58.

\item  M. Chaves and H. Morales, ``The Standard Model and the Generalized
Covariant Derivative'', in \emph{Proceedings of the International Workshop
``Lorentz Group, CPT, and Neutrinos''}, Universidad Aut\'{o}noma de Zacatecas,
M\'{e}xico, June 23-26, 1999, (World Scientific) 188.

\item  M. Chaves and H. Morales, \emph{Mod. Phys. Lett.} A, \textbf{15}, 197 (2000).
\end{enumerate}
\end{document}